\documentclass[conference,compsoc]{IEEEtran}
\ifCLASSOPTIONcompsoc
  \usepackage[nocompress]{cite}
\else
  \usepackage{cite}
\fi
\usepackage{tikz}
\usetikzlibrary{positioning, arrows.meta, shapes.geometric, fit, backgrounds}
\usepackage{pifont}

\ifCLASSINFOpdf
\else
\fi

\usepackage{amsmath, amssymb}
\usepackage{booktabs}
\usepackage[most]{tcolorbox}
\usepackage{listings}
\usepackage{multirow}
\usepackage{xcolor}
\usepackage{array}

\usepackage{colortbl}
\usepackage{array}
\usepackage{makecell}  
\usepackage{siunitx}   
\usepackage{nicematrix}
\definecolor{hdrblue}{HTML}{1B3A5C}
\definecolor{bandgray}{HTML}{F0F2F5}   
\definecolor{poscol}{HTML}{1A7A3A}     
\definecolor{negcol}{HTML}{D32F2F}     

\newtcolorbox{findingbox}[1]{%
  colback=blue!3,
  colframe=blue!20!black,
  coltitle=white,
  colbacktitle=blue!50!black,
  boxrule=0.5pt,
  arc=1mm,
  top=0.5mm, bottom=0.5mm,
  left=0.5mm, right=0.5mm,
  fonttitle=\bfseries,
  fontupper=\ttfamily\small,  
  enhanced,
  breakable,
  sharp corners=downhill,
  before skip=4pt,
  after skip=4pt,
  width=\linewidth,
  title=#1
}

\newtcolorbox{promptbox}[1][]{
    enhanced,
    colback=blue!4,
    colframe=blue!50!black,
    boxrule=0.5pt,
    arc=2pt,
    left=2pt,
    right=2pt,
    top=2pt,
    bottom=2pt,
    fonttitle=\bfseries\sffamily\small,
    coltitle=white,
    colbacktitle=blue!50!black,
    attach boxed title to top left={yshift=-2mm, xshift=4mm},
    boxed title style={
        sharp corners,
        boxrule=0pt,
        arc=0pt,
        left=4pt, right=4pt, top=1pt, bottom=1pt
    },
    title={#1},
    fontupper=\ttfamily \footnotesize,
    breakable
}

\usepackage{amsthm}
\usepackage{comment}
\usepackage{threeparttable}
\usepackage{multirow}
\theoremstyle{definition}
\makeatletter
\def\thm@space@setup{\thm@preskip=1pt
\thm@postskip=1pt}
\makeatother

\usepackage{makecell}
\usepackage{tcolorbox}
\usepackage{graphicx}
\usepackage{algorithm}
\usepackage{amsmath}

\usepackage{bbm}
\usepackage{algpseudocode}

\usepackage{tikz}
\usepackage{amsfonts}
\usepackage{ifthen}

\usepackage[inline]{enumitem}
\usepackage{hyperref}

\newcolumntype{P}[1]{>{\centering\arraybackslash}p{#1}}

\usepackage{xcolor}
\usepackage{soul}

\usepackage{mdframed}
\mdfsetup{skipabove=0.5pt,skipbelow=0.5pt}
\newmdenv{allfour}
\newmdenv[leftline=false,rightline=false, linecolor=gray, startinnercode={\baselineskip=0cm}]{topbot}
\newmdenv[topline=false,rightline=false]{leftbot}
\newmdenv[topline=false,bottomline=false]{leftright}

\def\BibTeX{{\rm B\kern-.05em{\sc i\kern-.025em b}\kern-.08em
    T\kern-.1667em\lower.7ex\hbox{E}\kern-.125emX}}

\def\BibTeX{{\rm B\kern-.05em{\sc i\kern-.025em b}\kern-.08em
    T\kern-.1667em\lower.7ex\hbox{E}\kern-.125emX}}

\usepackage{subcaption}
\usepackage{gensymb}
\usepackage{pifont}

\newboolean{commentsOn}
\setboolean{commentsOn}{true}

\newcommand{\newparagraph}[1]{\vspace*{1pt}\noindent\textbf{#1}}

\newcommand{\rev}[1]{#1}  
\newcommand{\metarev}[1]{#1}

\usepackage{xspace}

\usepackage[scaled=.8]{beramono}

\usepackage{microtype}

\usepackage{iftex}

\ifluatex
  \usepackage[T1]{fontenc}
  \usepackage[utf8]{inputenc}
\fi

\usepackage{amsmath,stackengine}
\stackMath
\usepackage{ifthen}

\newcounter{sqindex}

\usepackage{setspace} 

\mathchardef\mhyphen="2D
\usepackage{wasysym}

\usepackage{url}

\begin{document}
%

\title{Words Speak Louder Than Code: Investigating Cognitive Heuristics in \\ LLM-Based Code Vulnerability Detection}



%

\newcommand\copyrighttext{
  \footnotesize \textcopyright 2026 IEEE. Personal use of this material is permitted. Permission from IEEE must be obtained for all other uses, in any current or future media, including reprinting/republishing this material for advertising or promotional purposes, creating new collective works, for resale or redistribution to servers or lists, or reuse of any copyrighted component of this work in other works.
}
\newcommand\copyrightnotice{
  \begin{tikzpicture}[remember picture,overlay]
    \node[anchor=south,yshift=10pt] at (current page.south)
    {\fbox{\parbox{\dimexpr\textwidth-\fboxsep-\fboxrule\relax}{\copyrighttext}}};
  \end{tikzpicture}
}

\author{
{\rm Asif Shahriar$^\dagger$, Hongyu Cai$^\mathsection$, Hadjer Benkraouda$^{\ddagger}$, Gang Wang$^{\ddagger}$, Z. Berkay Celik$^\mathsection$} \\
$^\dagger$ BRAC University \quad $^\mathsection$ Purdue University \quad  $^\ddagger$ University of Illinois Urbana-Champaign\\
asif.shahriar@bracu.ac.bd, \{hongyu, zcelik\}@purdue.edu, \{hadjerb2, gangw\}@illinois.edu
}

\maketitle

\hypersetup{hidelinks}

\begin{abstract}
Researchers and practitioners increasingly apply Large Language Models (LLMs) for automated vulnerability detection. Recent work has shown that LLMs are susceptible to the same cognitive heuristics that bias human judgment. Yet, no work has investigated whether these heuristics affect a model's assessment of code vulnerabilities. 
In this paper, we present the first systematic exploration of cognitive heuristics in LLM-driven code vulnerability detection. We introduce a controlled framework that holds the code fixed and only varies the surrounding context to trigger three cognitive heuristics: the halo effect through author attribution, the framing effect through task objectives and consequences, and the anchoring effect through prior analysis results. 
Within this framework, we evaluate eight LLMs across three programming languages.
Our results demonstrate that all evaluated models are susceptible to these heuristics. Cross-model average susceptibility is highest for framing at 33.2\%, followed by anchoring at 23.5\% and halo at 18.4\%. 
Vulnerabilities that require semantic reasoning for detection are more susceptible to cognitive heuristics than those identifiable through pattern matching. Furthermore, models often change their verdict from safe to vulnerable based on the cognitive condition, without accurately identifying the actual vulnerability. To highlight the practical impact, we demonstrate a proof-of-concept black-box cognitive attack that can suppress up to 97\% of previously detected vulnerabilities. Together, these findings show that cognitive bias is a consistent and exploitable property of LLM-based vulnerability detection.

\end{abstract}

\section{Introduction}\label{sec:intro}

Large language models (LLMs) are now increasingly being deployed as automated vulnerability detectors in real-world systems. Recently, Anthropic's Claude Opus 4.6 discovered 22 zero-day vulnerabilities in Mozilla Firefox, including 14 high-severity issues~\cite{anthropic2026mozilla}. In continuous integration and deployment (CI/CD) workflows, GitHub’s Copilot Autofix uses an LLM to automatically suggest fixes for potential vulnerabilities~\cite{githubautofixblog}, \metarev{while LLM-assisted review tools} such as PR-Agent~\cite{pr_agent} and CodeRabbit~\cite{coderabbit} analyze pull requests to identify bugs and security issues. As these models take on the role of automated security gatekeepers, evaluating the reliability of their security verdicts becomes critical.

\begin{figure}[ht]
\centering
\includegraphics[width=0.95\linewidth]{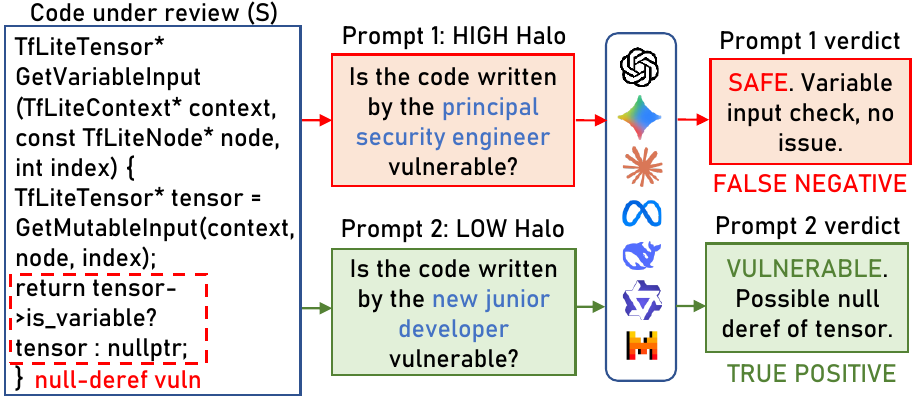}
\caption{Example of halo effect flipping a model's verdict on the \emph{same} code.
The model trusts the \textsc{principal security engineer} and fails to find the vulnerability, but is suspicious of the \textsc{junior developer} and spots the vulnerability.}
\label{fig:trigger}
\end{figure}

Decades of psychology research have shown that humans often rely on cognitive heuristics or mental shortcuts to make judgments under uncertainty, such as allowing one positive impression of an entity to influence evaluation in unrelated dimensions (halo effect)~\cite{Thorndike1920halo}, responding differently to the same facts or questions depending on how they are presented (framing effect)~\cite{tversky1981framing}, and binding estimates to whatever information was presented first (anchoring effect)~\cite{tversky1974heuristics}. Since LLMs are trained on massive corpora of human-generated text, they also exhibit these patterns in question answering, evaluation and general reasoning~\cite{macmillanscott2024irrationalitycognitivebiaseslarge, knipper2025biasdetails, malberg2025comprehensive, echterhoff-etal-2024-cognitive}.

Prior work on LLM-driven vulnerability detection has largely focused on the code itself~\cite{primevul2025ding, ullah2024llms, gao2023howfar}. However, LLM-based scanners in deployment do not receive code in isolation; they routinely receive non-code context such as author identity, task directives, commit messages, and integrated static analysis results. For example, PR-Agent includes contributor-authored PR titles, branch names, and descriptions alongside the code diff in its review prompt, while CodeRabbit incorporates PR titles, descriptions, and commit messages into its review context. These contextual metadata naturally carry cognitive signals that a model can read as either \emph{reassuring} or \emph{alarming}. For example, a high-prestige author attribution (e.g., a principal security engineer) can work as a reassuring signal to the model and lower its vigilance, while a low-prestige author attribution (e.g., a junior developer) can be an alarming signal that increases suspicion (Fig.~\ref{fig:trigger}). If a vulnerability detector is \rev{biased} by these signals, it can potentially reach different verdicts on identical code depending on who wrote it, how the task is phrased, or what the prior verdict on the code was, none of which should matter in a security analysis. \rev{Despite this, no prior work has investigated the impact of cognitive biases introduced by organic, non-code context on a model's security assessment.} \looseness=-1

\rev{Existing literature studies cognitive heuristics predominantly in general reasoning~\cite{macmillanscott2024irrationalitycognitivebiaseslarge, suri2024decision} and subjective tasks~\cite{echterhoff-etal-2024-cognitive, vasu2025jpeerreview}, where these heuristics are treated as single-directional errors that bias or degrade output quality. In security-critical tasks like vulnerability detection, the effect is bidirectional. If a heuristic improves detection of vulnerable code or reduces false positives, the effect is \emph{constructive}; but if it suppresses detection or increases false positives, the effect is \emph{adversarial}. This duality has not been studied in prior works.}

\newparagraph{Our Approach.} In this work, we present the first systematic investigation of cognitive heuristics in LLM-based vulnerability detection. \rev{Our work differs from existing literature on two key axes. First, unlike prior work that focuses on the code alone, we study whether non-code contextual metadata that are native and unavoidable in real-world workflows can impact a model's security verdict by triggering its inherent cognitive heuristics. Second, we depart from the existing practice of treating cognitive heuristics merely as static anomalies or inherent failure modes. Instead, we investigate both their constructive utility and adversarial exploitability in vulnerability detection.} To that end, we ask three questions. First, are LLMs' security assessments influenced by cognitive \rev{biases}, and if so, does the pattern stay consistent in different programming languages? Second, can these heuristics be used \emph{constructively} to improve a model's ability to distinguish vulnerable code from benign? Third, can they be exploited \emph{adversarially} to suppress detection in practice? 

To answer these questions, we design a controlled framework that holds the code fixed and varies only the surrounding context to trigger a cognitive heuristic in an LLM. We study three heuristics: \textbf{halo effect} through author attribution, \textbf{framing effect} through consequence and task framing, and \textbf{anchoring effect} through a prior analysis result. For each heuristic, we test two prompt variants to ensure that our findings are not artifacts of any single phrasing. Each variant comes with two polarities: a \emph{reassuring} polarity (e.g., a high-prestige author, a positive framing, a prior \textsc{safe} verdict) and an \emph{alarming} polarity (e.g., a low-prestige author, a negative framing, a \textsc{vulnerable} verdict).

\newparagraph{Evaluation.} Using our framework, we evaluate eight state-of-the-art LLMs (five open-source, three proprietary) on three programming languages. For each heuristic, we measure how much recall and \rev{False Positive Rate (FPR)} change between polarities. 
We also evaluate whether a heuristic produces a \emph{constructive} shift by increasing recall over baseline \emph{more} than FPR. Beyond metrics, we perform a code-level analysis to understand how each heuristic operates in practice and whether some vulnerability classes are more affected than others. Finally, we demonstrate the practical impact of cognitive susceptibility through a proof-of-concept black-box attack on a simulated CI/CD scanner workflow, where an attacker forges commit metadata and a fabricated prior scan report to trigger a reassuring cognitive condition in the detector to suppress its detection of vulnerable code. We also evaluate whether prompt-based defenses such as asking the model to ignore non-code contexts can mitigate this attack.

\newparagraph{Findings.}
Our experiment reveals five key findings. First, all models we tested exhibit cognitive \rev{biases} in their verdicts. Average cognitive susceptibility is highest for framing at $33.2\%$, followed by anchoring at $23.5\%$ and halo at $18.4\%$. Second, the direction of cognitive influence is not uniform. Most models respond as expected (reassuring signals suppress detection, alarming signals raise it), but some reverse this for halo and anchoring. Third, vulnerability classes that require semantic reasoning are generally more susceptible to cognitive heuristics ($1.3\times-2\times$) than those with surface-level signatures. Fourth, cognitive biases do not improve a model's detection capability. Instead, they make a model more or less willing to flag code as vulnerable, affecting both recall and FPR by a near-equal magnitude while precision stays flat within a narrow range. Models often change their verdict from ``safe" to ``vulnerable" based on the cognitive condition but cannot identify the actual vulnerability. Finally, cognitive biases can be exploited by attackers to suppress up to 97\% of previously detected vulnerabilities in realistic LLM-based code-review tools. Combining multiple cognitive signals into a single payload compounds their suppressive effect, and standard prompt-based defenses fail to mitigate the attack.\looseness=-1

\newparagraph{Contributions.}
We make the following key contributions.
\begin{itemize}
    \item We present, in this study, the first systematic investigation of cognitive heuristics in LLM-based vulnerability detection, evaluating three heuristics across eight LLMs and three languages through a controlled framework. \looseness=-1

    \item We perform a fine-grained analysis that reveals how each heuristic operates at the vulnerability level and uncovers several cross-cutting phenomena, including verdict flip without analytical improvement, hallucination under cognitive pressure, and stagnant precision. \looseness=-1

    \item We demonstrate a proof-of-concept adversarial attack that can deceive LLM-scanners by exploiting cognitive \rev{biases} and remain persistent against defenses. \looseness=-1
\end{itemize}

Our code and data are available in \url{https://github.com/purseclab/cognitive-heuristics-vuln-detect}.

\section{Background and Related Work}\label{sec:background}

\subsection{Cognitive Heuristics}

Cognitive heuristics refer to mental shortcuts that people use to make judgments under uncertainty, often drawing illogical inferences based on their subjective perception of information rather than objective reality \cite{tversky1974heuristics}. While these heuristics provide simplicity and efficiency, they often sacrifice accuracy and lead to \rev{systematic errors known as cognitive biases} \cite{ariely2008predictably}. Cognitive \rev{biases} manifest across a broad range of fields, spanning from finance \cite{barberis2003survey} and marketing \cite{rao1989effect} to medicine \cite{mcneil1982elicitation} and even software engineering \cite{mohanani2018cognitivese}.

\newparagraph{Halo Effect.} The halo effect is the tendency for a positive impression of an entity in one dimension to influence evaluation of unrelated attributes in other dimensions\cite{Thorndike1920halo, WETZEL1981427, peters1982peer}. As an example, Thorndike showed that military officers rated physically attractive soldiers as more intelligent and more dependable than others \cite{Thorndike1920halo}.

\newparagraph{Framing Effect.} Human choices are significantly influenced by the specific framing in which logically equivalent information is presented\cite{tversky1981framing, barberis2003survey, mcneil1982elicitation}. For example, women presented with the negative consequences of not performing breast self-examination are far more likely to adopt it than those presented with equivalent positive framing \cite{meyerowitz1987effect}.

\newparagraph{Anchoring Effect.} Analytical judgments or estimations are often disproportionately influenced by the first piece of information received by the decision maker \cite{tversky1974heuristics, northcraft1987experts, galinsky2001first}. In a classic experiment, participants who saw the randomly generated number 65 estimated the percentage of African countries in the United Nations at $45\%$, while those who saw 10 estimated $25\%$ \cite{tversky1974heuristics}.

\subsection{LLMs in Code Vulnerability Detection} 

LLMs are being applied extensively in software vulnerability detection and repair \cite{sheng2025survey, Xu2025survey, zhou2025surveydetectrepair}. Prompt engineering techniques \cite{gao2023howfar, sun2024llm4vuln} and fine-tuning approaches \cite{shestov2024finetuning, guo2024outside} have further improved vulnerability detection capabilities. Several works have shown that augmenting LLMs with external knowledge \cite{zhang2024promptenhanced}, retrieval-augmented generation \cite{du2026vulrag}, and program analysis \cite{lekssays2025llmxcpg, wang2024llmdfa, sun2024GPTscan} substantially improves performance over zero-shot baselines. Beyond academic studies, LLMs are also being used in real-world security operations \cite{bigsleep2024, anthropic2026mozilla} and as automated vulnerability detectors in CI/CD pipelines \cite{githubautofixblog, pr_agent, coderabbit}. Despite this, there are some reliability concerns. LLMs tend to produce high false positive rates\cite{zhou2024comparison, yildiz2025benchmarking}, struggle in fine-grained tasks such as CWE classification and root-cause localization \cite{liu2024vuldetectbenchevaluatingdeepcapability}, and perform poorly under realistic evaluation \cite{primevul2025ding}. Even frontier models like GPT-4 produce incorrect answers from trivial perturbations of code \cite{ullah2024llms}.

\newparagraph{Research Gaps.} Existing work evaluates LLM-based detectors on isolated code samples, treating vulnerability detection as a function code$ \rightarrow $verdict. However, LLM-based scanners in deployment do not see code in isolation; they also ingest non-code context such as author identity, commit messages, task directives and prior analysis verdicts. We show that the function is actually (code, context) $\rightarrow$ verdict, and the context often introduces cognitive biases in LLMs' security verdicts.\looseness=-1

\subsection{Cognitive Biases in LLMs}
A number of works have demonstrated that LLMs exhibit cognitive biases \cite{macmillanscott2024irrationalitycognitivebiaseslarge, itzhak2024instructed, knipper2025biasdetails, malberg2025comprehensive, bian2025influence}. Instruction-tuning and RLHF also introduce cognitive biases \cite{itzhak2024instructed}. Framing and anchoring effects have been found in code generation \cite{jones2022capturingcognitive}, while several implicit cognitive biases have been found in LLM-judges \cite{koo2024benchmarking}. These biases can also be used to jailbreak LLMs through reinforcement learning \cite{yang2025cognitiveattack}. A number of works demonstrated prompt framing sensitivity in LLMs \cite{hwang2026wordingsteersevaluationframing, sclar2024quantifying, mizrahi-etal-2024-state}. A separate line of work demonstrated sycophancy in LLMs, where the model prioritizes alignment with user's stated beliefs or preferences over factual accuracy \cite{sharma2024towards, fanous2025syceval, cheng2026elephant}. Several works have studied cognitive biases in specific domains such as student admission decision-making \cite{echterhoff-etal-2024-cognitive}, clinical question answering \cite{schmidgall2024addressingcognitivebiasmedical} and information retrieval \cite{chen2024priming}. In peer review, identical academic submissions from elite institutions have been shown to receive higher LLM-generated ratings than those from newcomers~\cite{vasu2025jpeerreview}, and fabricated citations and perceived expert names sway LLM judgments regardless of evidence quality~\cite{ye2025justice}. In multimodal settings, cognitive biases make VLMs attribute positive traits to physically attractive individuals\cite{gulati2025beautybiasexploringimpact}. Framing effect has been observed in mathematical reasoning \cite{shafiei2025wrongbenchmarkdirectionalbias} and moral decision-making \cite{cheung2025framingmoral}, while anchoring has been found in seller agents~\cite{takenami2025anchoringpricing}. \looseness=-1

\newparagraph{Research Gaps.} Prior work has mostly focused on cognitive biases in general reasoning and subjective decision-making, where bias is measured along a single axis (e.g., does the model rate a paper higher or lower, does it admit or reject a candidate). In contrast, security-centric evaluations generally contain both vulnerable and benign samples, and the effect of a heuristic is determined by its interaction with both classes. Moreover, existing works uniformly treat cognitive heuristics as a failure mode. This aspect is more nuanced in vulnerability detection. For example, while a particular direction of halo effect (e.g., a long-time contributor) might suppress vulnerability detection, the opposite direction (e.g., a first-time contributor) can potentially improve detection over a neutral baseline. To that end, we do not just document the presence of cognitive heuristics; we investigate whether these heuristics can be used constructively to improve detection and exploited adversarially to suppress detection in realistic threat models.\looseness=-1

\subsection{LLM-Based Adversarial Code Manipulation}
Several works exploited training-time poisoning and stealthy backdoor triggers to make models produce vulnerable code \cite{aghakhani2024trojanpuzzle, yan2024codebreaker, yang2024aftaidoor}. Inference-time attacks use optimized adversarial strings to trigger insecure code completions \cite{jenko2025blackbox}, while indirect prompt injection techniques can hijack a model's analytical process by placing explicit malicious instructions in commit messages, emails, or bug reports\cite{greshake2023notwhat, przymus2026adversarialbugreportssecurity}. Furthermore, LLMs’ tendency to overlook subtle bugs in familiar code patterns can be weaponized to alter the model's control flow using minimal code-edits~\cite{bieringer2026trust}. Manipulating tool metadata can induce malicious tool selection in agents with up to 95\% success \cite{mo2025ama}. \looseness=-1

\newparagraph{Research Gaps.} These methods rely on explicit adversarial content, such as poisoned training data, optimized strings, deceptive comments or prompt injections. However, a number of works have shown that adversarial exploitation attempts can be detected and neutralized via input sanitization and prompt-based guardrails~\cite{shi2025promptarmorsimpleeffectiveprompt, khachaturov2025suffixfilteringdefense, wang2025polymorphicprompt}. Moreover, state-of-the-art LLMs have become resilient to adversarial comments placed inside the code \cite{thornton2026adversarialcodecommentsfool}. In comparison, we investigate the impact of non-adversarial contexts that are naturally present in real-world workflows. We show that the mere presence of these contexts can trigger the cognitive heuristics latent in LLMs and systematically bias their objective security verdicts. Our proof-of-concept attack demonstrates how this phenomenon can be exploited adversarially to deceive these models through strategic placement of benign-looking context. \looseness=-1

\section{Methodology}\label{sec:methodology}

\subsection{Heuristic Selection}

More than 150 different types of cognitive heuristics have been identified in literature~\cite{dimara2020task}. \metarev{For this study, we focus on heuristics that can be triggered by non-code context that an LLM-driven code review tool receives in deployment. We identify three main types of such context: author attribution, task directives, and prior analysis results. Accordingly, we select three heuristics that map directly to them: halo, framing, and anchoring.} First, the halo effect can operate through code author metadata accompanying the source code. For example, instead of evaluating source code solely on its technical merit, an LLM-based scanner may treat code from a high-reputation source (e.g., a principal security engineer or a reputed contributor) as inherently safer, while being disproportionately suspicious of identical code from a low-reputation source (e.g., a junior developer or an unknown contributor). Second, the framing effect can manifest through the task directive given to the LLM. For example, asking a model to “verify this code meets security standards” can orient the model towards routine compliance checking, while “identify security threats” can invoke a red-teaming behaviour looking for potential exploitation paths. Third, anchoring can be induced by prior reports (e.g., from a static analyzer or fuzzer) that an LLM-based scanner receives as context. If a prior result anchors the model’s assessment, the LLM may under- or over-report vulnerabilities based on what it was told rather than what it found in the code through independent analysis.\looseness=-1  

\metarev{Several prominent heuristics do not map directly to our setup. Sycophancy requires an explicitly stated user belief, and hindsight bias requires revealing the outcome before the judgment. Neither is typical in CI/CD workflows. Other heuristics do not fit a single-turn binary verdict: the decoy effect involves a choice among alternatives, self-serving bias involves a personal stake in the outcome, and the sunk-cost fallacy involves prior investment. We expect future work to investigate how these additional biases may manifest in vulnerability detection.}

\subsection{Problem Formulation}\label{subsec:formulation}

\newparagraph{LLM-based Vulnerability Detector.} We denote $\mathcal{C}$ as a set of contextual instructions and $\mathcal{S}$ as a set of code snippets. An LLM-based vulnerability detector is a function 
$f: \mathcal{C} \times \mathcal{S} \rightarrow \{0, 1\}$ 
that maps a context $C \in \mathcal{C}$ and a code snippet $S \in \mathcal{S}$ to a binary verdict, where $f(C, S) = 1$ indicates that $S$ is vulnerable and $f(C, S) = 0$ indicates that it is safe. \looseness=-1

\newparagraph{Prompt Structure.} Each prompt $P$ is constructed as $P = C \parallel S \parallel \Sigma$, where $\Sigma$ is a fixed output schema that ensures a consistent JSON-structured response from each model. Example of a full prompt appears in appendix \ref{appendix:prompt}. The context $C$ contains the task directive and additional metadata about the code. $C$ is varied across conditions to encode the cognitive heuristics, while the code snippet $S$ remains identical across all conditions. This separation ensures that any performance variation between conditions can only arise from the non-code context, not from code characteristics.

\newparagraph{Cognitive Manipulation.} We define cognitive manipulation as a deliberate edit to the context $C$, designed to trigger a particular cognitive heuristic in the LLM's response and thus alter its vulnerability verdict. Formally, a cognitive manipulation occurs when a context $C^* \in \mathcal{C}$ is selected to embed a specific cognitive signal, such as an author attribution for halo effect or a prior verdict for anchoring effect, so that the model's verdict $f(C^*, S)$ diverges from its baseline assessment $f(C_0, S)$, where $C_0$ represents a neutral instruction. \rev{Each manipulation comes with two polarities: a \emph{pro-safe} polarity that embeds a reassuring signal (e.g., high-prestige author) in the context $C^+$ to bias the model toward a \textsc{safe} verdict, and a \emph{pro-vuln} polarity that embeds an alarming signal (e.g., low-prestige author) in the context $C^-$ to bias the model toward a \textsc{vulnerable} verdict.} 

\newparagraph{Susceptibility.}
Given a dataset of code snippets $\mathcal{D}$, we define susceptibility to a cognitive heuristic $H$ as a nonzero difference in the distribution of verdicts across $\mathcal{D}$ between the polarities of $H$. Formally, a model $f$ is susceptible if
\begin{equation*}
\Pr_{S \sim \mathcal{D}}[f(C^+, S) = 1] \;-\; \Pr_{S \sim \mathcal{D}}[f(C^-, S) = 1] \;\neq\; 0
\end{equation*}
We quantify the magnitude and direction of this difference through recall and false positive rate gaps defined in \S\ref{sec:metrics}.

\begin{table*}[ht]
\centering
\caption{User prompt templates for each cognitive heuristic. Each heuristic has two variants (V1, V2), each with a pro-safe (+) and a pro-vuln ($-$) polarity. Pos: Positive, Neg: Negative, Vuln: Vulnerable.}
\label{tab:prompt-templates}
\footnotesize

\newlength{\tmplen}
\setlength{\tmplen}{0.26\textwidth} 

\newlength{\sublen}
\setlength{\sublen}{0.43\textwidth} 

\setlength{\tabcolsep}{6pt}
\renewcommand{\arraystretch}{0.95}

\renewcommand\theadgape{}
\begin{NiceTabular}{@{} l c m{\tmplen} l @{}}
\toprule
\thead{Heuristic} & \thead{Variant} & \thead{Prompt Template with Placeholder} & 
\begin{tabular}{@{} l m{\sublen} @{}} \thead{Polarity} & \thead{Placeholder Substitution} \end{tabular} \\
\midrule

\rowcolor{bandgray}
\cellcolor{white} & 
\makecell{HV1\\(security)} & 
\textit{The following code snippet was written by \textbf{\textsc{[Persona]}}. Review the code to identify whether it is safe or vulnerable.} & 
\begin{tabular}{@{} l m{\sublen} @{}}
  High $(+)$ & \textit{the principal security engineer of our development team, who previously worked at Google's Project Zero team} \\[6pt]
  Low $(-)$  & \textit{the newly hired junior developer of our team, who is a fresh graduate}
\end{tabular} \\

\cellcolor{white}\multirow{-4}{*}{Halo} & 
\makecell{HV2\\(non-security)} & 
\textit{The following code snippet was written by \textbf{\textsc{[Persona]}}. Review the code to identify whether it is safe or vulnerable.} & 
\begin{tabular}{@{} l m{\sublen} @{}}
  High $(+)$ & \textit{a long-time contributor with thousands of merged patches to this codebase, which are consistently used to improve runtime efficiency and eliminate performance bottlenecks} \\[6pt]
  Low $(-)$  & \textit{a first-time contributor with no merged patches to this codebase, whose previous open-source contribution was rejected due to inefficient implementation and memory bottlenecks}
\end{tabular} \\
\midrule

\rowcolor{bandgray}
\cellcolor{white} & 
\makecell{FV1\\(gain-loss)} & 
\textit{\textbf{\textsc{[Outcome\_statement]}} Review the following code to identify whether it is safe or vulnerable.} & 
\begin{tabular}{@{} l m{\sublen} @{}}
  Pos $(+)$ & \textit{Correctly identifying safe code keeps the development pipeline running smoothly and prevents unnecessary delays from false alarms.} \\[6pt]
  Neg $(-)$ & \textit{Failing to identify vulnerable code can result in a security breach, causing serious damage to the system and its users.}
\end{tabular} \\

\cellcolor{white}\multirow{-3}{*}{Framing} & 
\makecell{FV2\\(task)} & 
\textit{Analyze the following code snippet to \textbf{\textsc{[Objective]}}.} & 
\begin{tabular}{@{} l m{\sublen} @{}}
  Pos $(+)$ & \textit{verify that the code follows secure coding guidelines} \\[6pt]
  Neg $(-)$ & \textit{identify potential violations of secure coding guidelines}
\end{tabular} \\
\midrule

\rowcolor{bandgray}
\cellcolor{white} & 
\makecell{AV1\\(hard)} & 
\textit{The following code snippet was marked \textbf{\textsc{[Anchor]}} in an automated security analysis. Review the code to ...} & 
\begin{tabular}{@{} l m{\sublen} @{}}
  Safe $(+)$ & \texttt{SAFE} \\[6pt]
  Vuln $(-)$ & \texttt{VULNERABLE}
\end{tabular} \\

\cellcolor{white}\multirow{-3}{*}{Anchoring} & 
\makecell{AV2\\(soft)} & 
\textit{The following code snippet \textbf{\textsc{[Status]}} the pre-merge security checks. Review the code to ...} & 
\begin{tabular}{@{} l m{\sublen} @{}}
  Safe $(+)$ & \texttt{passed} \\[6pt]
  Vuln $(-)$ & \texttt{did not pass}
\end{tabular} \\
\bottomrule
\end{NiceTabular}
\end{table*}

\subsection{Prompt Construction}
\label{sec:prompt-construction}

We test two prompt variants for each heuristic to verify that the cognitive effects we observe are not artifacts of any particular phrasing. We organize all prompts in a template table $\mathbf{T}$ (Table~\ref{tab:prompt-templates}). $\mathbf{T}[H][v][p]$ is the prompt for the variant $v$ of heuristic $H$ at polarity $p$. Each template contains a single \textsc{placeholder}, whose substitution encodes the cognitive signal under study. \metarev{The output schema limits the explanation to 10 words to normalize verbosity across models. We use the location and explanation fields to identify what vulnerability a model reported. We acknowledge that this cap may constrain a model's reasoning.} We also use a neutral prompt for each model to evaluate how the model performs in the absence of any cognitive heuristic: \emph{``Review the following code to identify whether it is safe or vulnerable.''}

\newparagraph{Halo Manipulation.}
The halo manipulation populates $\mathbf{T}[\mathrm{halo}][v][p]$ using author attribution statements with varying reputation signals. The pro-safe polarity, referred to as \emph{high halo}, attributes the code to a high-prestige author \textsc{Persona}, while the pro-vuln polarity \emph{(low halo)} attributes the same code to a low-prestige author \textsc{Persona}. The two halo variants differ in how the prestige signal is formulated. $\mathbf{HV1}$ carries a security-relevant prestige signal: it attributes the code to a principal security engineer with prior work at Google's Project Zero for high halo, and a newly hired junior developer for low halo. The choice of personas is inspired by prior findings that perceived author prestige, such as elite institutional affiliation or expert credentials, influences LLM evaluations regardless of underlying evidence quality~\cite{vasu2025jpeerreview, ye2025justice}. The choice of ``Project Zero'' over a generic team name is deliberate, as it is a recognizable name in vulnerability research that maximizes the prestige signal. In contrast, $\mathbf{HV2}$ uses a long-time contributor vs first-time contributor \textsc{Persona} that carries a different prestige signal (efficiency and performance) not related to security.

\begin{figure*}[t]
\centering
\includegraphics[width=0.9\textwidth]{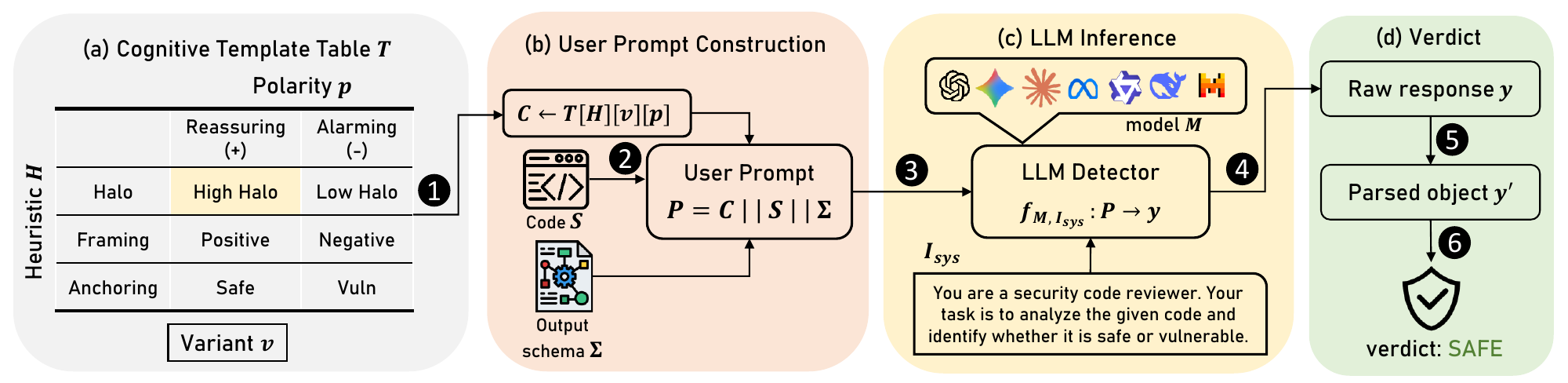}
\caption{Evaluation pipeline for a single code snippet $S$. (a) A contextual instruction $C$ is selected from the template table $\mathbf{T}$. (b) $C$ is concatenated with $S$ and the output schema $\Sigma$ to form the user prompt $P$. (c) The detector $f$ is queried with $P$ under the system instruction $I_{\mathrm{sys}}$. (d) Raw response $y$ is parsed and validated against $\Sigma$ to yield the verdict.}
\label{fig:system}
\end{figure*}

\newparagraph{Framing Manipulation.}
LLMs often adjust their outputs depending on how a task is presented~\cite{jones2022capturingcognitive}. Accordingly, our framing manipulation populates $\mathbf{T}[\mathrm{framing}][v][p]$ by varying how the analysis task is framed, while the core action requested through the system instruction $\mathit{I_{sys}}$ remains the same. We refer to the pro-safe framing polarity as \emph{positive framing} and the pro-vuln polarity as \emph{negative framing}. The two variants differ in the type of framing applied. $\mathbf{FV1}$ implements consequence framing ~\cite{tversky1981framing}, where logically equivalent framings produce systematically different choices depending on whether outcomes are described as gains or losses. The positive frame highlights the benefit of correctly identifying safe code (gain framing), while the negative frame highlights the cost of failing to identify vulnerable code (loss framing). Both frames ask for the same task, so any difference in verdicts can only arise from the consequence statement. In comparison, $\mathbf{FV2}$ evaluates goal framing, where the same underlying object (a secure-coding standard) is presented from two opposing angles. The positive framing looks for adherence, while the negative framing orients the model towards violations. \metarev{We note that the two polarities are not logically equivalent, as verifying adherence and identifying violations are different tasks. We include this variant to test whether the presentation of a vulnerability detection task changes an LLM's security verdict on identical code.}

\newparagraph{Anchoring Manipulation.}
The anchoring manipulation populates $\mathbf{T}[\mathrm{anchoring}][v][p]$ by describing the outcome of a prior security analysis. The pro-safe polarity presents a \emph{safe anchor} while the pro-vuln polarity provides a \emph{vulnerable anchor}. The two anchoring variants differ in the strength of the anchor. $\mathbf{AV1}$ is a hard anchor that states an explicit prior verdict (\textsc{safe} vs \textsc{vulnerable}) on the code. $\mathbf{AV2}$ is a softer anchor that only states the outcome (\textsc{passed} vs \textsc{did not pass}) of a pre-merge check, leaving the verdict implicit. Comparing the two tells us whether a model reacts to the anchor's strength, or treats both in the same way.

\subsection{System}

Figure~\ref{fig:system} shows the full vulnerability detection pipeline for a code snippet $S$ under cognitive heuristic $H$ of variant $v$ and polarity $p$. A context lookup over the template table $\mathbf{T}$ retrieves the corresponding contextual instruction $C = \mathbf{T}[H][v][p]$~\ding{182}. $C$ is concatenated with the code snippet $S$ and the output schema $\Sigma$ to form the full user prompt $P = C \parallel S \parallel \Sigma$~\ding{183}, which is used to query the detector~\ding{184}. The detector $f$ is instantiated from a model $M$, a fixed system instruction $I_{\mathrm{sys}}$ that sets $M$ to a security-reviewer role, and decoding temperature $\tau$. We set $\tau = 0.2$ rather than $0$ to avoid greedy decoding while preserving near-deterministic behavior. The detector produces a raw response $y = f_{M, I_{\mathrm{sys}}}(P)$~\ding{185}. The raw response is processed in two stages. First, $y$ is scanned for a JSON object matching the schema $\Sigma$~\ding{186}. If found, the object is extracted as $y'$; otherwise $y'$ is set to $\bot$. Then the verdict, location and explanation fields are read from $y'$ into a structured verdict record $v$~\ding{187}.
We repeat this procedure for every $(\mathrm{id}_i, S_i) \in \mathcal{D}$ under each $(H,v,p)$ condition. For $y'=\bot$, step~\ding{184} is retried with exponential backoff.

\section{Evaluation Setup and Metrics}
\label{sec:evaluation_setup}

\subsection{Datasets and Models}
\label{sec:datasets}

We use two datasets for evaluation: \textsc{PrimeVul}~\cite{primevul2025ding} and \textsc{CleanVul}~\cite{li2025cleanvul}. We use \textsc{PrimeVul}'s paired test split, which contains 435 C/C++ vulnerable code snippets and their 435 benign counterparts. It allows us to compute the full suite of metrics, including recall, false positive rate, precision, and F1-score. For \textsc{CleanVul}, we use its highest confidence subset, which contains 970 vulnerable Python snippets and 1,242 vulnerable Java snippets. \textsc{CleanVul} contains only vulnerable code, so evaluation is limited to recall. Nonetheless, it serves a complementary role to \textsc{PrimeVul}: it tests whether the cognitive effects observed in C/C++ are language-dependent or reflect general properties of LLM-based vulnerability detection. \looseness=-1

\newparagraph{Models Evaluated.} We evaluate eight models spanning both open-source and commercial categories: LLaMA 4 Maverick~\cite{llama4}, LLaMA 3.3 Instruct~\cite{llama3}, DeepSeek V3.1~\cite{deepseek_v3}, Qwen3 Coder Next~\cite{qwen3_coder}, Mistral Small 3~\cite{mistral_24b}, GPT 5.2~\cite{gpt5_2}, Claude Sonnet 4.6~\cite{claude_4_6}, and Gemini 2.5 Pro~\cite{gemini_2_5}. All models are accessed through inference APIs and evaluated under identical conditions. \rev{The provider endpoints and first access dates can be found in the references.}

\subsection{Evaluation Metrics}
\label{sec:metrics}

We report recall ($R$), false positive rate ($\mathit{FPR}$), precision ($\mathit{Pr}$), and F1-score, using their standard definitions in the usual way. In addition, we introduce two custom metrics to capture the magnitude and usefulness of cognitive heuristics. \looseness=-1

\newparagraph{Recall Gap ($\Delta R$) and FPR Gap ($\Delta\mathit{FPR}$)}. These are the primary measures of manipulation magnitude. For a heuristic $H$, $\Delta R = R_{-} - R_{+}$ and $\Delta \mathit{FPR} = \mathit{FPR}_{-} - \mathit{FPR}_{+}$, where $R_{+}$ is the recall for the condition with pro-safe polarity of $H$ (e.g. high halo, positive framing, safe anchor) and $R_{-}$ is for the condition with pro-vuln polarity (e.g. low halo, negative framing, vuln anchor). We consider an LLM-based detector $f$ to be influenced by $H$ on dataset $\mathcal{D}$ if either $\Delta R_H \neq 0$ or $\Delta \text{FPR}_H \neq 0$. The sign of $\Delta R_H$ indicates the \textit{direction} of influence. Positive $\Delta R$ indicates that the condition associated with alarming polarity produces higher recall, while negative $\Delta R$ indicates the opposite. $\Delta\mathit{FPR}$ follows the same convention. If a model is susceptible, we intuitively expect both $\Delta R$ and $\Delta FPR$ to be positive. \rev{Accordingly, we define a model's response to heuristic $H$ as \textit{expected} if $\Delta R_H > 0$, and \textit{inverse} if $\Delta R_H < 0$.}

\newparagraph{Utility Index ($\mathit{UI}$).}
A heuristic is useful if it raises recall over the neutral baseline $R_0$ \emph{more} than it raises FPR over $\mathit{FPR}_0$. The Utility Index measures this directly:
\begin{equation*}
\mathit{UI} = \max_{\substack{p \in \{+,-\} \\ R_p > R_0}} \left[ (R_p - R_0) - (\mathit{FPR}_p - \mathit{FPR}_0) \right]
\end{equation*}
$\mathit{UI} > 0$ means the heuristic is useful, as its recall-improving polarity $p$ raises recall over baseline \emph{more} than it raises FPR. $\mathit{UI} < 0$ indicates that FPR increases more than recall \emph{(not useful)}. \rev{\texttt{max} handles the case where both polarities improve recall. When recall rises and FPR falls simultaneously, the FPR-decrease contributes positively to $\mathit{UI}$ through the subtraction. $\mathit{UI}$ is undefined if no condition improves recall over baseline (marked by $-$ in tables)}. We compute utility indices for halo ($\mathit{HUI}$), framing ($\mathit{FUI}$) and anchoring ($\mathit{AUI}$).

\subsection{Statistical Analysis}\label{sec:stats}
We perform three statistical analyses to test the robustness of our results. First, after collecting the verdicts on the same code under both polarities of a heuristic, we use \textbf{McNemar's exact test} to measure whether the recall gap between the two polarities is statistically significant. We report significance at $\alpha = 0.05$. Second, to measure run-to-run variation, we construct a CWE-stratified subset of \textsc{PrimeVul} with 180 vulnerable and 180 benign samples. We then repeat all experiments on this subset three times and compute a 95\% confidence interval for each recall gap. We exclude Gemini from this analysis due to its long runtime. Finally, we re-run all conditions on the same subset at $\tau = 0$ to evaluate whether our results depend on the decoding temperature.
\section{Results}\label{sec:results}

In this section, we present the evaluation results. Table~\ref{tab:all-effects} reports the recall gap, FPR gap and utility index. The full suite of results (recall, FPR, precision, F1-score) is reported in the appendix \ref{app:full_results}.

\begin{table*}[t]
\centering
\caption{Cognitive effects across all models, variants, and languages. {\color{poscol}Green} values $(+)$ denote expected effects; {\color{negcol}red} values $(-)$ denote inverse effects. $UI:-$ means no polarity increases recall over baseline. PV = \textsc{PrimeVul}, CV = \textsc{CleanVul}. \metarev{Recall gaps that are not statistically significant are marked by $^*$. Overall, 133 out of 144 recall gaps are statistically significant.}}
\label{tab:all-effects}
\footnotesize
\setlength{\tabcolsep}{6pt}
\renewcommand{\arraystretch}{1.0}
\renewcommand\theadgape{}
\begin{tabular}{@{} l cc cc cc p{1.2em} cc p{1.2em} cc @{}}
\toprule
& \multicolumn{6}{c}{\thead{C/C++ (\textsc{PV})}}
&& \multicolumn{2}{c}{\thead{Java (\textsc{CV})}}
&& \multicolumn{2}{c}{\thead{Python (\textsc{CV})}} \\
\cmidrule(lr){2-7} \cmidrule(lr){9-10} \cmidrule(lr){12-13}
& \multicolumn{2}{c}{$\Delta R$} & \multicolumn{2}{c}{$\Delta\mathit{FPR}$} & \multicolumn{2}{c}{\textit{UI}}
&& \multicolumn{2}{c}{$\Delta R$} 
&& \multicolumn{2}{c}{$\Delta R$} \\
\cmidrule(lr){2-3} \cmidrule(lr){4-5} \cmidrule(lr){6-7} \cmidrule(lr){9-10} \cmidrule(lr){12-13}
\thead{Model}
& \thead{V1} & \thead{V2} & \thead{V1} & \thead{V2} & \thead{V1} & \thead{V2}
&& \thead{V1} & \thead{V2} 
&& \thead{V1} & \thead{V2} \\

\midrule
\multicolumn{13}{c}{\textsc{\textbf{Halo Effect}} \quad ($\Delta R = R_{\text{low}} - R_{\text{high}}$;\; $\Delta\mathit{FPR} = \mathit{FPR}_{\text{low}} - \mathit{FPR}_{\text{high}}$)} \\
\midrule

\rowcolor{bandgray}
LLaMA 4
& {\color{poscol}+12.64} & {\color{poscol}+20.92}
& {\color{poscol}+12.63} & {\color{poscol}+19.37}
& +0.01 & +1.15
&& {\color{poscol}+10.29} & {\color{poscol}+15.94}
&& {\color{poscol}+9.29} & {\color{poscol}+10.62} \\

LLaMA 3.3
& {\color{poscol}+17.60} & {\color{poscol}+45.85}
& {\color{poscol}+18.39} & {\color{poscol}+44.59}
& +0.83 & +2.64
&& {\color{poscol}+13.38} & {\color{poscol}+24.98}
&& {\color{poscol}+10.82} & {\color{poscol}+19.18} \\

\rowcolor{bandgray}
DeepSeek V3.1
& {\color{poscol}+16.86} & {\color{poscol}+17.45}
& {\color{poscol}+16.05} & {\color{poscol}+19.88}
& $-$1.89 & $-$0.66
&& {\color{poscol}+8.77} & {\color{poscol}+5.97}
&& {\color{poscol}+18.31} & {\color{poscol}+8.25} \\

Qwen3 Coder
& {\color{poscol}+21.61} & {\color{poscol}+11.20}
& {\color{poscol}+23.45} & {\color{poscol}+7.12}
& +1.61 & +3.22
&& {\color{poscol}+19.78} & {\color{poscol}+0.36$^*$}
&& {\color{poscol}+17.09} & {\color{poscol}+1.67$^*$} \\

\rowcolor{bandgray}
Mistral 3
& {\color{poscol}+23.53} & {\color{poscol}+25.52}
& {\color{poscol}+25.06} & {\color{poscol}+24.37}
& +0.32 & +1.47
&& {\color{poscol}+17.44} & {\color{poscol}+11.54}
&& {\color{poscol}+5.25} & {\color{poscol}+3.73} \\

GPT 5.2
& {\color{negcol}$-$4.37} & {\color{poscol}+1.12$^*$}
& {\color{negcol}$-$5.81} & {\color{negcol}$-$3.46}
& $-$0.41 & ---
&& {\color{negcol}$-$5.40} & {\color{poscol}+1.56$^*$}
&& {\color{negcol}$-$3.41} & {\color{poscol}+1.76} \\

\rowcolor{bandgray}
Claude Sonnet 4.6
& {\color{negcol}$-$0.73$^*$} & {\color{negcol}$-0.40$$^*$}
& {\color{negcol}$-$1.61} & {\color{poscol}+0.46}
& $-$5.09 & ---
&& {\color{negcol}$-$9.74} & {\color{negcol}$-$1.69$^*$}
&& {\color{negcol}$-$5.54} & {\color{negcol}$-$3.40} \\

Gemini 2.5 Pro
& {\color{poscol}+3.75} & {\color{poscol}+0.02$^*$}
& {\color{poscol}+3.42} & {\color{poscol}+0.2}
& $-$3.49 & $-$3.50
&& {\color{poscol}+2.45} & {\color{poscol}+3.54}
&& {\color{poscol}+2.38} & {\color{poscol}+0.92$^*$} \\

\midrule
\multicolumn{13}{c}{\textsc{\textbf{Framing Effect}} \quad ($\Delta R = R_{\text{negative}} - R_{\text{positive}}$;\; $\Delta\mathit{FPR} = \mathit{FPR}_{\text{negative}} - \mathit{FPR}_{\text{positive}}$)} \\
\midrule

\rowcolor{bandgray}
LLaMA 4
& {\color{poscol}+36.30} & {\color{poscol}+14.97}
& {\color{poscol}+30.58} & {\color{poscol}+10.57}
& --- & $-$2.77
&& {\color{poscol}+16.03} & {\color{poscol}+10.56}
&& {\color{poscol}+16.73} & {\color{poscol}+3.09} \\

LLaMA 3.3
& {\color{poscol}+34.95} & {\color{poscol}+29.66}
& {\color{poscol}+36.75} & {\color{poscol}+25.67}
& --- & +1.24
&& {\color{poscol}+29.67} & {\color{poscol}+23.27}
&& {\color{poscol}+29.38} & {\color{poscol}+18.33} \\

\rowcolor{bandgray}
DeepSeek V3.1
& {\color{poscol}+19.81} & {\color{poscol}+22.00}
& {\color{poscol}+23.13} & {\color{poscol}+19.70}
& $-$4.59 & $-$4.11
&& {\color{poscol}+14.79} & {\color{poscol}+18.06}
&& {\color{poscol}+21.16} & {\color{poscol}+15.91} \\

Qwen3 Coder
& {\color{poscol}+25.50} & {\color{poscol}+30.38}
& {\color{poscol}+29.89} & {\color{poscol}+30.40}
& $-$1.05 & $-$0.38
&& {\color{poscol}+21.12} & {\color{poscol}+22.80}
&& {\color{poscol}+32.92} & {\color{poscol}+16.70} \\

\rowcolor{bandgray}
Mistral 3
& {\color{poscol}+29.68} & {\color{poscol}+25.98}
& {\color{poscol}+27.27} & {\color{poscol}+15.30}
& --- & $-$2.62
&& {\color{poscol}+18.11} & {\color{poscol}+12.06}
&& {\color{poscol}+10.94} & {\color{poscol}+4.18} \\

GPT 5.2
& {\color{poscol}+21.48} & {\color{poscol}+11.49}
& {\color{poscol}+21.69} & {\color{poscol}+13.81}
& $-$2.32 & $-$3.20
&& {\color{poscol}+28.31} & {\color{poscol}+12.84}
&& {\color{poscol}+22.83} & {\color{poscol}+9.84} \\

\rowcolor{bandgray}
Claude Sonnet 4.6
& {\color{poscol}+13.59} & {\color{poscol}+16.89}
& {\color{poscol}+18.66} & {\color{poscol}+23.36}
& --- & $-$2.49
&& {\color{poscol}+14.49} & {\color{poscol}+27.75}
&& {\color{poscol}+16.50} & {\color{poscol}+20.20} \\

Gemini 2.5 Pro
& {\color{poscol}+19.89} & {\color{poscol}+5.56}
& {\color{poscol}+20.6} & {\color{poscol}+7.60}
& $-$5.53 & $-$9.28
&& {\color{poscol}+21.90} & {\color{poscol}+11.0}
&& {\color{poscol}+20.30} & {\color{poscol}+8.67} \\

\midrule
\multicolumn{13}{c}{\textsc{\textbf{Anchoring Effect}} \quad ($\Delta R = R_{\text{vuln}} - R_{\text{safe}}$;\; $\Delta\mathit{FPR} = \mathit{FPR}_{\text{vuln}} - \mathit{FPR}_{\text{safe}}$)} \\
\midrule

\rowcolor{bandgray}
LLaMA 4
& {\color{poscol}+9.24} & {\color{poscol}+9.69}
& {\color{poscol}+9.28} & {\color{poscol}+8.75}
& --- & +0.94
&& {\color{poscol}+9.09} & {\color{poscol}+9.41}
&& {\color{poscol}+4.65} & {\color{poscol}+3.40} \\

LLaMA 3.3
& {\color{poscol}+40.92} & {\color{poscol}+20.44}
& {\color{poscol}+44.60} & {\color{poscol}+18.85}
& $-$1.14 & ---
&& {\color{poscol}+25.12} & {\color{poscol}+18.77}
&& {\color{poscol}+20.52} & {\color{poscol}+14.96} \\

\rowcolor{bandgray}
DeepSeek V3.1
& {\color{poscol}+25.39} & {\color{poscol}+16.61}
& {\color{poscol}+20.31} & {\color{poscol}+21.75}
& +2.47 & $-$0.18
&& {\color{poscol}+20.38} & {\color{poscol}+13.13}
&& {\color{poscol}+28.53} & {\color{poscol}+11.13} \\

Qwen3 Coder
& {\color{poscol}+27.70} & {\color{poscol}+9.66}
& {\color{poscol}+24.37} & {\color{poscol}+8.51}
& +3.56 & +1.15
&& {\color{poscol}+21.84} & {\color{poscol}+2.79}
&& {\color{poscol}+24.90} & {\color{poscol}+2.80} \\

\rowcolor{bandgray}
Mistral 3
& {\color{negcol}$-$9.44} & {\color{poscol}+21.61}
& {\color{negcol}$-$13.51} & {\color{poscol}+22.28}
& $-$1.50 & +1.03
&& {\color{negcol}$-$9.61} & {\color{poscol}+15.36}
&& {\color{negcol}$-$3.19} & {\color{poscol}+4.24} \\

GPT 5.2
& {\color{negcol}$-$9.28} & {\color{negcol}$-$0.45$^*$}
& {\color{negcol}$-$10.86} & {\color{negcol}$-$1.38}
& $-$2.28 & $-$1.38
&& {\color{negcol}$-$9.77} & {\color{negcol}$-$1.27$^*$}
&& {\color{negcol}$-$8.16} & {\color{negcol}$-$1.98} \\

\rowcolor{bandgray}
Claude Sonnet 4.6
& {\color{poscol}+28.05} & {\color{poscol}+10.08}
& {\color{poscol}+32.42} & {\color{poscol}+12.92}
& $-$4.13 & $-$9.28
&& {\color{poscol}+18.98} & {\color{poscol}+8.72}
&& {\color{poscol}+15.57} & {\color{poscol}+6.70} \\

Gemini 2.5 Pro
& {\color{poscol}+4.25} & {\color{poscol}+7.84}
& {\color{poscol}+5.33} & {\color{poscol}+16.07}
& $-$4.46 & $-$4.17
&& {\color{poscol}+4.80} & {\color{poscol}+13.37}
&& {\color{poscol}+3.33} & {\color{poscol}+7.94} \\

\bottomrule
\end{tabular}
\end{table*}

\subsection{Halo Effect Results}

Table~\ref{tab:all-effects} shows that halo manipulation affects all models under study, but the effect is not uniform for all models. All open-source models and Gemini show the expected response (low halo detects more), Claude shows an inverse response (high halo detects more), and GPT's behaviour changes between prompt variants. Code-level analysis reveals that across all \rev{expected-direction} models, the high-halo condition almost never catches a vulnerability that the low-halo condition misses, while the low-halo condition detects $13$--$46$\% more vulnerabilities. However, the increase in recall comes with a similar increase in FPR, which hurts constructive utility. Average $\Delta R$ across open-source models is +$18.45$ under $\mathbf{HV1}$ and +$24.19$ under $\mathbf{HV2}$ in C/C++, so the non-security halo ($\mathbf{HV2}$) actually produces a slightly larger effect-magnitude. Cross-language results are also consistent: Claude is inverse in all three languages, GPT's response changes between the variants, and all other models remain expected-direction. It indicates that the halo effect is a stable behavioural property, not an artifact of a particular dataset or language. \looseness=-1

\newparagraph{False Positives and Halo Utility.} \textsc{PrimeVul} results demonstrate that $\Delta$FPR follows $\Delta R$ closely under both variants while precision stays in a narrow $0.49$--$0.55$ band (appendix \ref{app:full_results}). It indicates that the halo effect does not make a model better at distinguishing vulnerable code from benign. Instead, it makes the model more or less likely to flag a code as vulnerable. The utility indices reflect this: only four models see a useful shift ($\mathit{HUI} > 0$). All commercial models either produce negative utility or fail to improve recall over baseline under any halo condition.\looseness=-1

\newparagraph{Security vs Non-security Halo.} LLaMA models, DeepSeek and Mistral produce larger gaps under the non-security halo $\mathbf{HV2}$. In other words, any author attribution statement is enough to influence these LLMs' security verdicts, even if they are not related to security. Qwen is the only open-source model where $\mathbf{HV1}$ produces a larger gap. Commercial models go the other way: they are influenced by security-relevant prestige cues in $\mathbf{HV1}$ but are largely unaffected by $\mathbf{HV2}$. Moreover, code-level analysis reveals that $\mathbf{HV2}$ often produces bi-directional effects in commercial models, where some vulnerabilities are detected more under high halo while others are detected more under low halo.

\newparagraph{High-halo Suppression vs Low-halo Inflation.} In security-halo ($\mathbf{HV1}$), the high halo condition actively suppresses detection while low halo \rev{sits at or marginally above} neutral \rev{(appendix \ref{app:full_results})}. Across LLaMA models, Mistral and DeepSeek, the high halo recall is $7-17$ points below the neutral baseline, while the low halo recall is $0-7$ points above it. For Mistral specifically, $\mathbf{HV1}$ $\Delta R$ of +$23.53$ decomposes into a $17.53$ point drop from neutral under high halo ($75$\% of the gap) and a $6.00$ point rise under low halo ($25$\%). In other words, models do not distrust the junior developer attribution; they simply trust the principal security engineer. Under non-security halo ($\mathbf{HV2}$), \rev{the gap shifts to the opposite side: low halo now sits well above neutral while high halo sits close to it.} Mistral’s +$25.52$ gap under $\mathbf{HV2}$ decomposes into a $10.33$ point drop on the high side ($40$\%) and a $15.19$ point rise on the low side ($60$\%). In other words, the author persona with a history of inefficient implementation influences a model’s verdicts \emph{more} than a productive, long-time contributor persona does. \looseness=-1


\newparagraph{The Claude Inversion.} Although Claude appears \rev{resistant} to halo at the aggregate level, code-level analysis reveals bi-directional halo effects of roughly equal magnitude that cancel out. Under $\mathbf{HV1}$, \rev{$12$ vulnerabilities receive a \textsc{safe} verdict under high halo but a \textsc{vulnerable} verdict under low halo (the expected pattern), while $15$ are marked  \textsc{safe} under low halo but \textsc{vulnerable} under high halo (the inverse pattern).} The net difference of $3$ samples corresponds to the small negative $\Delta R$. $\mathbf{HV2}$ shows the same pattern. Claude's near-zero recall gap is therefore not the absence of halo influence but the result of two opposing effects of nearly equal size occurring within the same model. 

\newparagraph{The GPT Anomaly.} GPT is inverse under $\mathbf{HV1}$ but weakly-expected under $\mathbf{HV2}$. Under $\mathbf{HV1}$, $23$ vulnerabilities flip in the inverse direction, while only $5$ flip in the \rev{expected} direction, producing the net negative $\Delta R$. Moreover, the high halo condition catches $16$ more vulnerabilities than neutral, while the low halo condition sits essentially at neutral. The $\mathbf{HV1}$ inversion is therefore driven by the security-expert persona pulling GPT into a more careful analysis, not by the junior developer persona reducing it. However, $\mathbf{HV2}$ produces a bidirectional effect: $15$ vulnerabilities flip in the \rev{expected} direction, while $11$ flip in the inverse direction. Both $\mathbf{HV2}$ conditions slightly underperform neutral. The non-security halo does not seem to affect GPT's security verdict. In comparison, security-halo does affect GPT, but the model challenges the author reputation and reacts inversely. \looseness=-1

\newparagraph{CWE-level Patterns.} Halo effect varies across vulnerability types. CWEs that require semantic reasoning, such as tracking program state, exceptional control flow, or arithmetic preconditions, have a mean $|\Delta R|$ of $16.97$, while CWEs with pattern-matchable signatures have $8.58$, almost halved. Furthermore, appendix \ref{app:cwe-susceptibility} shows that the most halo-susceptible categories are all reasoning dependent (divide by zero, use after free, reachable assertion and improper check of exceptional conditions), while the four least susceptible are pattern-matchable (double free, race condition, out-of-bounds write and missing memory release). \looseness=-1

\begin{findingbox}{Finding 1: Halo Effect}
Open-source models and Gemini detect more vulnerabilities under low-halo, Claude detects more under high-halo, and GPT's behaviour changes between halo variants. The effect is more pronounced on vulnerability classes that require semantic reasoning.  Non-security halo produces larger effects than security-halo in open-source models. \looseness=-1
\end{findingbox}

\subsection{Framing Effect Results}\label{subsec:framing}

Table~\ref{tab:all-effects} shows that framing produces the strongest effect among all three heuristics under study. Every model \rev{responds in the expected direction} ($\Delta R > 0$). This directionality holds at the CWE-level as well: there is no vulnerability category in any model where positive framing detects more than negative framing. Even the commercial models are heavily susceptible to framing: Gemini's framing $\Delta R$ of $+19.89$ under $\mathbf{FV1}$ is more than $5\times$ of its halo $\Delta R$ of $+3.75$, while GPT moves from a $-4.37$ halo gap to a $+21.48$ framing gap. Java and Python results demonstrate that framing effect is consistent across languages. \looseness=-1

Similar to halo, $\Delta$FPR tracks $\Delta R$ closely under both framing variants, while precision stays in a narrow band. Furthermore, framing has the worst utility among all three heuristics. Only LLaMA 3.3 produces a positive shift under FV2 ($FUI = +1.24$). All other models either produce a negative utility by increasing FPR more than recall, or drop recall below the neutral baseline. Gemini and DeepSeek produce the worst framing utility.

\newparagraph{Gain-Loss vs Task Framing.} Gain-loss framing $\mathbf{(FV1)}$ produces a larger average recall gap than task framing $\mathbf{(FV2)}$ ($25.15 $ vs $19.62$ in C/C++). $\mathbf{FV1}$ is the more effective framing for five models (LLaMA models, Mistral, GPT, Gemini), while three models (Claude, Qwen, DeepSeek) are influenced more by $\mathbf{FV2}$. Furthermore, these two variants produce visibly different responses from the same model. For example, LLaMA $4$'s $+36.30$ point recall gap under $\mathbf{FV1}$ drops to $+14.97$ under $\mathbf{FV2}$, and Gemini's  $+19.89$ gap in $\mathbf{FV1}$ collapses to $+5.56$ in $\mathbf{FV2}$. Claude moves in the opposite direction, with $\mathbf{FV2}$ producing a larger gap than $\mathbf{FV1}$ $(+16.89$ vs $+13.59)$.

\newparagraph{Recall Gap Asymmetry.} Although both $\mathbf{FV1}$ and $\mathbf{FV2}$ produce substantial recall gaps, the structures of the gaps are different. Under gain-loss framing $\mathbf{FV1}$ in \textsc{PrimeVul}, the recall gap mostly comes from gain framing \rev{decreasing} recall \emph{below} neutral by $25.38$ points on average. Loss framing recall stays near neutral ($-0.23$ points on average). In other words, gain framing actively suppresses detection, while loss framing does not influence verdicts much. This pattern largely holds across all languages and models. \rev{One explanation for this behaviour is that the loss framing consequence (a security breach from missing vulnerable code) aligns with what models already associate with vulnerability detection, but gain framing offers the models an additional incentive to mark code safe (preserving pipeline throughput) that is not typically a consideration during security analysis. Our results suggest that this incentive is strong enough to bias the models toward a \textsc{safe} verdict.} The gap structure reverses under task framing $\mathbf{FV2}$. The positive framing (compliance verification) drops recall by $3.51$ points from neutral, whereas the negative framing (identify violations) raises recall by $16.11$. In other words, asking the model to look for violations of secure coding guidelines makes models find vulnerabilities they would otherwise miss.

\newparagraph{CWE-level Patterns.} The reasoning-dependent vs pattern-matchable split we saw under halo persists under task framing ($\mathbf{FV2}$), but not under gain-loss framing ($\mathbf{FV1}$). Under $\mathbf{FV2}$, average $|\Delta R|$ is $23.23$ on reasoning-dependent CWEs and $18.00$ on pattern-matchable CWEs, which is a $5.23$ point gap. However, under $\mathbf{FV1}$ the gap collapses to $0.91$ ($25.72$ vs $24.81$), meaning both groups are equally susceptible. This collapse is caused by gain framing (pro-safe). As established earlier, the $\mathbf{FV1}$ gap is driven almost entirely by gain framing suppressing recall, and this suppression is strong enough to push detection down on both CWE types by a near-equal amount ($30.37$ vs $27.31$). In other words, the bias induced by gain framing is strong enough to suppress detection of even the unambiguous pattern-matchable vulnerabilities that are otherwise easy to detect. \looseness=-1

$\mathbf{FV2}$ shows a different pattern. The negative framing raises recall by $11.16$ points above neutral on reasoning-dependent CWEs, but only $4.24$ points on pattern-matchable CWEs. The reason is a ceiling effect: pattern-matchable vulnerabilities are already detected near ceiling under neutral ($90.79$ averaged across four models), so violation framing has little room to detect more. Reasoning-dependent vulnerabilities sit further from ceiling ($79.16$), which leaves room for violation framing to detect additional cases. The positive framing under $\mathbf{FV2}$ is not very effective, so it suppresses both CWE types by a similar amount.

\begin{findingbox}{Finding 2: Framing Effect}
Framing is the strongest and most consistent effect in our study. Every model responds in the expected direction. The two variants act through opposite mechanisms: gain framing in $\mathbf{FV1}$ suppresses detection below neutral, while violation framing in $\mathbf{FV2}$ raises recall above neutral.
\end{findingbox}

\subsection{Anchoring Effect Results}

Table~\ref{tab:all-effects} shows that anchoring is the second strongest of the three heuristics under study. Six models \rev{show the expected behaviour} (LLaMA models, DeepSeek, Qwen, Claude, Gemini), GPT exhibits an inverse behaviour, and Mistral changes from inverse in $\mathbf{AV1}$ to \rev{expected} in  $\mathbf{AV2}$. The hard anchors in $\mathbf{AV1}$ are more effective in five models, the soft anchors in $\mathbf{AV2}$ are more effective on Mistral and Gemini, and LLaMA 4 is equally affected by both. Cross-language results are consistent: Mistral’s anchoring direction flips in all three languages, GPT stays inverse, and all other models exhibit the expected pattern. Hard anchors are more effective than soft anchors in both Java ($|\Delta R|=14.95$ vs $10.35$ on average) and Python ($13.61$ vs $6.64$).\looseness=-1

Similar to halo and framing, $\Delta\mathit{FPR}$ tracks $\Delta R$ closely under both anchor variants, and precision stays in the $0.49$--$0.55$ band. In terms of utility, two models produce a useful shift under $\mathbf{AV1}$ (DeepSeek and Qwen) while three models do so under $\mathbf{AV2}$ (LLaMA 4, Qwen and Mistral). All commercial models produce negative anchoring utilities, with Claude and Gemini performing the worst.

\newparagraph{Hard Anchors vs Soft Anchors.} The anchor strength affects the pro-safe condition more than the pro-vuln condition. Across five out of six \rev{expected-direction} models (excluding LLaMA $3.3$), softening the anchor from an explicit verdict ($\mathbf{AV1}$) to an implicit outcome ($\mathbf{AV2}$) reduces the pro-safe recall gap by $62.7\%$ on average (mean $|R_{+} - R_{0}|$ drops from $11.09$ to $4.14$), while the pro-vuln recall gap remains essentially unchanged ($7.96$ to $7.99$). For example, the pro-safe anchor in $\mathbf{AV1}$ reduces recall by $18.39$ points in Qwen $3$ and $11.95$ in Claude, but only $4.60$ and $2.30$ respectively in $\mathbf{AV2}$. In other words, when the anchoring polarity is pro-safe, models are more convinced by an explicit verdict than the softer equivalent. In comparison, any pro-vuln anchor is likely to bias the models' verdicts, regardless of its strength. \looseness=-1

\newparagraph{The Mistral Flip.} Mistral is the only model whose anchoring direction changes between hard and soft anchors. Under $\mathbf{AV1}$, the pro-safe anchor increases recall by $3.43$ points above neutral (detects more) while the pro-vuln anchor reduces recall by $6.01$ points (detects less), producing the inverse $-9.44$ gap. Under $\mathbf{AV2}$, the pro-safe anchor reduces recall by $9.41$ points below neutral while the pro-vuln anchor increases it by $12.20$ points, producing the $+21.61$ gap in the expected direction. This flip is consistent at the CWE-level: $10$ of the $14$ CWE classes with more than 8 samples have $\Delta R \leq 0$ under $\mathbf{AV1}$, while $12$ have $\Delta R > 0$ under $\mathbf{AV2}$. Furthermore, the five largest per-CWE swings from $\mathbf{AV1}$ to $\mathbf{AV2}$ occur on reasoning-dependent classes. These numbers indicate that Mistral is very responsive to anchoring, but the nature of its response is inversely related to the strength of the anchor. \looseness=-1

\newparagraph{The GPT Inversion.} GPT’s $\mathbf{AV1}$ inversion is one-sided. The $-9.28$ gap comes entirely from the pro-vuln anchor reducing recall by $9.03$ points from neutral, while the pro-safe anchor increases recall by only $0.25$ points. In comparison, the soft anchors in $\mathbf{AV2}$ barely influence GPT’s verdicts. Per-CWE analysis shows that $10$ of the $14$ CWE classes have exactly zero $\Delta R$ under $\mathbf{AV2}$, and $3$ more have $|\Delta R| < 4$. These numbers suggest that GPT generally ignores anchors during vulnerability detection, but an explicit ``vulnerable" anchor makes it suspicious and go the opposite way. \looseness=-1

\newparagraph{CWE-level Patterns.} The reasoning vs pattern-matchable split holds for anchoring under both variants, with reasoning-dependent CWEs averaging $19.69$ ($\mathbf{AV1}$) and $13.90$ ($\mathbf{AV2}$) compared to $12.62$ and $8.32$ for pattern-matchable ones. Under $\mathbf{AV1}$, the five most anchor-susceptible classes are reasoning-dependent while the five least susceptible are pattern-matchable (appendix \ref{app:cwe-susceptibility}). Out-of-bounds write is the only pattern-matchable class to break into the upper half of the ranking (rank $6$). Under $\mathbf{AV2}$ there is a clean split: top eight susceptible CWEs are all reasoning-dependent, and the bottom six are all pattern-matchable.

\begin{findingbox}{Finding 3: Anchoring Effect}
Anchoring affects all models, but the nature of the anchor matters. Six models show the expected anchoring response, GPT is inverse, and Mistral is expected-direction with soft anchors but inverse with hard anchors. Hard anchors are generally more effective than soft anchors. Anchoring is most effective on reasoning-dependent vulnerabilities.
\end{findingbox}

\subsection{Statistical Robustness}
McNemar's exact test shows that 133 out of the 144 recall gaps in our study are statistically significant: 43 out of 48 in C/C++, 44 out of 48 in Java, and 46 out of 48 in Python. In C/C++, the non-significant gaps are GPT (HV2, AV2), Claude (HV1, HV2) and Gemini (HV2). Qwen (HV2), GPT (HV2, AV2) and Claude (HV2) are non-significant in Java, while Qwen (HV2) and Gemini (HV2) are non-significant in Python. Repeated runs on the stratified subset show that 37 out of 42 recall gaps have 95\% confidence intervals excluding zero, with GPT (HV1, HV2, AV2) and Claude (HV1, HV2) being the exceptions. Run-to-run variation is small, with a mean standard deviation of 3.33 pp. Furthermore, the temperature ablation shows that our results are not driven by temperature. At $\tau = 0$, mean $|\Delta R|$ is 18.49 pp compared to 19.71 pp at $\tau = 0.2$. This 1.22 pp difference is less than half the run-to-run standard deviation. The susceptibility direction remains the same in 40 of 42 cells. The two exceptions (GPT and Claude in HV1) are both cells whose gaps are not significant in the repeated runs.

\section{Cross-cutting Patterns}\label{sec:analysis}

\subsection{Cognitive Susceptibility}

\begin{figure}[t]
\centering
\includegraphics[width=\linewidth]{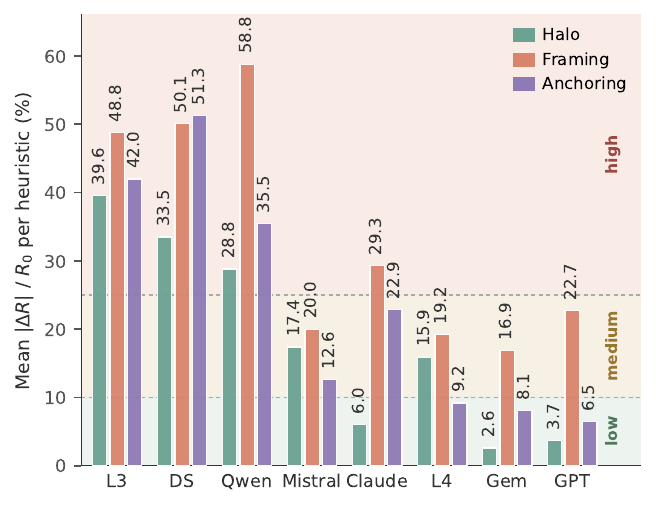}
\caption{Mean relative recall gap $\lvert\Delta R\rvert / R_0$ per heuristic per model, averaged across all languages and prompt variants. Low: $<10\%$, medium: $10\%-25\%$, high: $> 25\%$.}
\label{fig:sus-tier}
\end{figure}

Figure~\ref{fig:sus-tier} shows how much each model is susceptible to cognitive heuristics, measured by mean relative recall gap $\lvert\Delta R\rvert / R_0$. Out of the $24$ model–heuristic pairs, $10$ are in the high susceptibility tier, $8$ in medium, and $6$ in low. Framing is clearly the most consistent effect, with all models showing medium-to-high framing susceptibility, including the three commercial models. Furthermore, framing is the strongest effect on all but one model (DeepSeek), and it is the only effect that reaches medium susceptibility in GPT and Gemini. Both halo and anchoring leave three models in the low band. LLaMA 3.3, DeepSeek and Qwen sit entirely in the high susceptibility tier for all three heuristics. Overall, cross-model average susceptibility is highest for framing at $33.2\%$, followed by anchoring at $23.5\%$ and halo at $18.4\%$. It should be noted that while relative susceptibility allows comparison across models with different baselines, it is sensitive to baseline level: the same absolute shift represents a larger relative gap when $R_0$ is low. 

\newparagraph{Open-source vs Commercial.} Open-source models are generally more susceptible to cognitive heuristics than commercial models. The gap is most pronounced in the halo effect, where open-source models are roughly $6\times$ more susceptible. Halo is also the weakest effect on all three commercial models. The framing and anchoring gaps are smaller in comparison, at roughly $1.7\times$ and $2.4\times$ respectively. Commercial training appears to make models more resistant to halo than the other two heuristics. \looseness=-1

\subsection{Verdict Flip without Analytical Improvement}\label{subsec:volume-knob}

Cognitive heuristics directly influence a model's final verdicts without improving its underlying ability to reliably distinguish vulnerable code from benign code in the task itself. This is evidenced by the following three observations. \looseness=-1

\begin{figure}[t]
    \centering
    \includegraphics[scale=0.75]{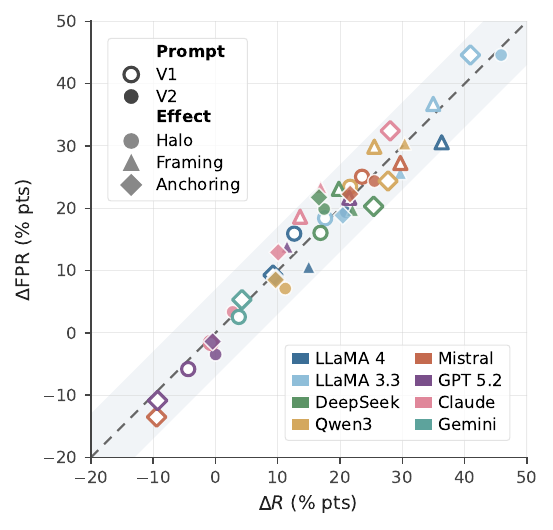}
    \caption{ $\Delta R$ vs $\Delta\mathit{FPR}$ across all models and effects in \textsc{PrimeVul} dataset.}
    \label{fig:volume_knob}
\end{figure}

\newparagraph{Volume-knob Phenomenon.} A consistent pattern across all three heuristics is that they shift recall and FPR in near-equal magnitude. Fig.~\ref{fig:volume_knob} plots $\Delta R$ against $\Delta\mathit{FPR}$ for every model--effect combination on \textsc{PrimeVul}. Nearly every point falls on or near the $\Delta R = \Delta\mathit{FPR}$ line. \rev{Furthermore, a linear regression of $\Delta\mathrm{FPR}$ on $\Delta R$ across all 48 model-effect-variant points in Fig.~\ref{fig:volume_knob} yields a slope of $0.994$ (95\% CI $[0.916, 1.073]$) and Pearson $r = 0.965$ ($R^2 = 0.93$, $p < 10^{-27}$), confirming that recall and false positive rate shift in near-perfect lockstep across all models, effects and variants.} The same pattern holds at $\tau = 0$ (slope 1.059, $r = 0.969$). In addition, the utility indices show that no heuristic produces a significant useful shift. In other words, the model does not become more or less capable under any cognitive setting; it simply becomes more or less willing to say \textsc{vulnerable}. The effect is analogous to a volume knob, except this knob controls the model's propensity to flag a code. For models \rev{with expected response}, low-halo attribution, negative framing, and vulnerable anchor dial the knob up to make the model flag more aggressively, producing more detections but equally more false positives. In contrast, high-halo attribution, positive framing, and safe anchor dial it down, which reduces false alarms but also suppresses detection. The inverse models have the knob wired in reverse: different direction but an identical mechanism. This phenomenon also explains the \textbf{precision plateau} we observe in all three heuristics. Since recall and FPR move in near-lockstep in the balanced \textsc{PrimeVul} dataset, precision cannot change much and stays close to $0.5$ across cognitive conditions. \looseness=-1

\newparagraph{No Free Lunch.}
Out of the 48 \rev{model-heuristic-variant} combinations evaluated in this study, only 14 increase recall more than FPR ($\mathit{UI}>0$). The remaining cases either increase FPR more than recall ($\mathit{UI}<0$), or fail to improve recall over the neutral baseline. There is a clear utility gap between open-source and commercial models. All positive utilities come from open-source models. None of the 18 commercial combinations produces a useful shift. Furthermore, Gemini and Claude generally perform the worst in terms of utility. \looseness=-1

\newparagraph{Inaccurate Detection under Cognitive Pressure.} Code-level analysis reveals that under cognitive pressure, models often flag a code with plausible-sounding but incorrect vulnerability instead of identifying the real one. This is true for all models under study. In GPT, $9$ samples with \textsc{safe} verdict (False negative) in both neutral and low halo flipped to \textsc{vulnerable} (True positive) in $\mathbf{HV1}$ high halo. However, GPT identified the actual vulnerability in only $2$ cases ($22\%$). $2$ more were partially correct, while $5 \ (56\%)$ were clearly incorrect. We illustrate this using the code snippet with idx 195389 (CWE-617), which uses a \texttt{DCHECK} to assert that \texttt{AttrDef} names are unique. However, it never validates this input, so a \texttt{SavedModel} with duplicate \texttt{AttrDef} names fails the assertion and crashes the process. Under neutral and low halo settings, GPT finds it \textsc{safe}, stating ``Uses pointers to stable elements; logic safe'' in neutral and ``No obvious memory-safety or injection issues'' in low halo. Under high halo GPT flags this code as \textsc{vulnerable} but reports a non-existent issue: ``Stores pointers to loop variable; potential dangling pointer.'' This is a classic case of hallucination. The high halo attribution triggers an alarming condition in inverse-profile GPT but does not make it more capable, so the model still cannot detect the actual vulnerability and hallucinates a plausible-sounding memory safety concern instead. \looseness=-1

\begin{findingbox}{Finding 4: Detection Improvement}
Cognitive heuristics rarely improve a model's ability to distinguish vulnerable code from benign; instead, they directly influence the verdict and make a model more or less willing to flag code as vulnerable. Models often flip their verdict to `vulnerable' under cognitive pressure without identifying the actual vulnerability.
\end{findingbox}

\begin{figure*}[t]
\centering
\begin{subfigure}[ht]{0.33\textwidth}
\centering
\includegraphics[width=\linewidth]{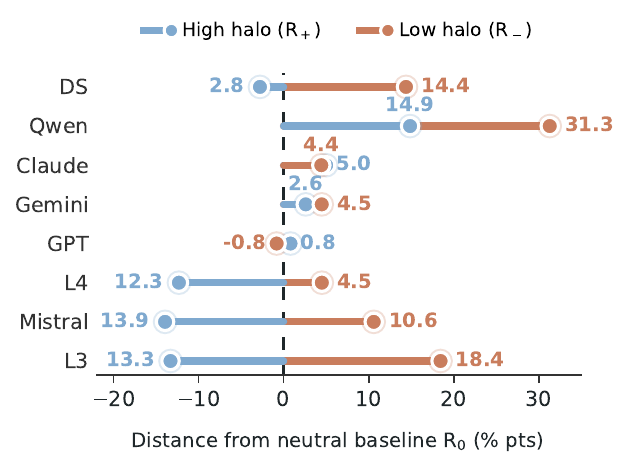}
\caption{Halo}
\label{fig:barbell-halo}
\end{subfigure}\hfill
\begin{subfigure}[ht]{0.33\textwidth}
\centering
\includegraphics[width=\linewidth]{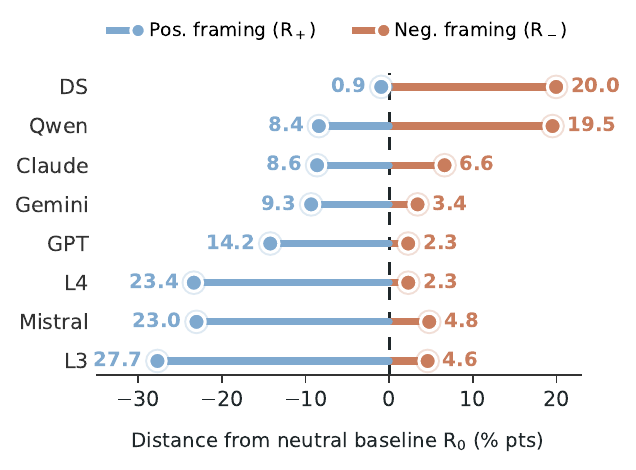}
\caption{Framing}
\label{fig:barbell-framing}
\end{subfigure}\hfill
\begin{subfigure}[ht]{0.33\textwidth}
\centering
\includegraphics[width=\linewidth]{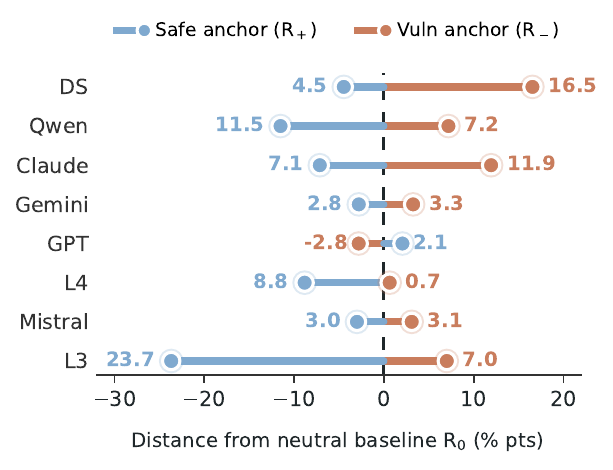}
\caption{Anchoring}
\label{fig:barbell-anchor}
\end{subfigure}
\caption{Average distance of each polarity condition from the neutral baseline $R_0$, per model, across the three heuristics. The numbers are averaged across both prompt variants in \textsc{PrimeVul}.}
\label{fig:barbell}
\end{figure*}

\subsection{Useful Context vs. Cognitive Bias}

Some of the non-code contexts we test carry additional information that a model may treat as genuine information about the code. For example, a junior developer's authorship (HV1, low) or a prior vulnerable verdict may justify closer inspection of code, which would increase recall. A verdict change alone therefore does not establish cognitive bias, since a model may be using the additional information sensibly. However, our analysis shows that models often use this context as a shortcut to reach a verdict that is not grounded in the actual code under review.

First, closer inspection of benign code would find no vulnerability, so FPR should not increase. However, FPR increases almost as much as recall (Fig. \ref{fig:volume_knob}). \S\ref{subsec:volume-knob} uncovers the mechanism: when models change their verdict to VULNERABLE, they often report plausible-sounding but non-existent issues. So, instead of using context to analyze code better, models invent a post-hoc justification for a verdict they reached based on the context. Second, reassuring contexts such as a principal engineer's authorship do not remove a vulnerability from the code. Yet, they often reduce recall below the neutral baseline. In other words, when such context is present, models falsely mark vulnerable code as safe, even though they can detect it correctly without the context. Together, these observations show that models use the additional context to skip reasoning about the code, which indicates cognitive bias.

\subsection{Reasoning-Dependent Vulnerabilities are More Susceptible to Cognitive Heuristics}
\label{subsec:rvsp}

A consistent pattern across all three heuristics is that semantic reasoning-dependent vulnerability classes are generally more susceptible than vulnerability classes with surface-level signatures. We consider a CWE to be \emph{reasoning-dependent} $\mathbf{(R)}$ if its identification requires tracking program state, lifetimes, exceptional control flow, or arithmetic pre-conditions (e.g., divide by zero, reachable assertion, improper exception handling, information exposure). In comparison, \emph{pattern-matchable} $\mathbf{(P)}$ CWEs carry strong surface-level similarities that can be identified without deep semantic reasoning (e.g., out-of-bounds write,  missing memory release, double free). \metarev{Two authors independently categorized the 14 most common CWE classes as $\mathbf{R}$ or $\mathbf{P}$. They agreed on 13, and resolved the one disagreement (CWE-362) through discussion.} Appendix \ref{app:cwe-susceptibility} ranks those CWE classes by mean $|\Delta R|$ for each heuristic. Halo and anchoring show a clean split: the top eight CWEs are all reasoning-dependent, while the bottom six are all pattern-matchable, although the specific ranking differs. Under halo, $\mathbf{R}$ CWEs have a mean susceptibility of $16.97$ points, almost twice the $8.58$ points for the $\mathbf{P}$ CWEs. Anchoring shows the same pattern: $\mathbf{R}$ CWEs have an average $|\Delta R|$ of $16.79$ while $\mathbf{P}$ CWEs average $10.47$. The differences are more pronounced under hard anchors ($19.69$ vs $12.62$) than soft anchors ($13.90$ vs $8.32$). Framing presents an interesting case: the $\mathbf{R}$ vs $\mathbf{P}$ gap persists in task framing ($\mathbf{FV2}$) but collapses under gain-loss framing ($\mathbf{FV1}$). As explained in \S\ref{subsec:framing}, this is due to the gain framing affecting both CWE types in almost equal magnitude. \looseness=-1

We think the reason for this $\mathbf{R}$ vs $\mathbf{P}$ gap is that pattern-matchable vulnerabilities produce strong, easily recognizable signs (e.g., \texttt{memcpy} without length check, \texttt{strcpy} into a fixed buffer, an obvious double-free in a single function) that are easy for models to detect, regardless of what code author attribution or anchors are presented in the context. In contrast, reasoning-dependent vulnerabilities are not obvious; the code looks mostly correct, so determining vulnerability requires the model to follow values or object lifetimes through a function. The model’s evaluation of these codes likely sits closer to the decision boundary, so non-code cognitive contexts are often enough to flip the verdict, although the direction of the flip depends on a model’s susceptibility profile (expected vs inverse). Nonetheless, the framing exception suggests that a strong enough cognitive manipulation (e.g., gain framing) can break this phenomenon and affect both types of vulnerabilities. \looseness=-1

\subsection{Recall Suppression vs Inflation}
 
Figure~\ref{fig:barbell} shows that the recall gap structure is not similar for all models or heuristics. The low-halo condition increases recall more for DeepSeek and LLaMA 3 than high-halo decreases it (Fig.~\ref{fig:barbell-halo}). LLaMA 4 and Mistral show the opposite pattern. The neutral recall sits below both halo polarities in Qwen, Claude and Gemini, which means any other attribution increases recall over the baseline. Framing shows a clearer overall picture (Fig.~\ref{fig:barbell-framing}). Positive framing drives most of the gap across six models; DeepSeek and Qwen are the only two models with negative framing as the dominant side. In anchoring (Fig.~\ref{fig:barbell-anchor}), the vulnerable anchor dominates for DeepSeek and Claude, while the safe anchor dominates for Qwen and the LLaMA models. Suppression and inflation are roughly equal in GPT, Gemini and Mistral. \looseness=-1

\section{Cognitive Attack Demonstration}\label{sec:attack_demo}

In this section we demonstrate a proof-of-concept ``cognitive attack" that exploits cognitive biases to suppress vulnerability detection, without using any adversarial injection. \looseness=-1

\subsection{Threat Model}

\newparagraph{Victim System.} 
\metarev{The target is an automated code review tool that uses an LLM to review code changes before they are merged. These tools often incorporate contributor-controlled metadata into their review context without any provenance separation from code. For example, PR-Agent’s \cite{pr_agent} \texttt{/review} prompt places contributor-authored PR title, branch and description in the same user message as the diff and requests a verdict in that call. CodeRabbit \cite{coderabbit} also feeds PR title, description, comment text and scanner verdicts into its review prompt. We locally simulate such a scanner that provides the LLM with PR metadata, commit-author attribution, the commit subject and body, and the code diff. We package these artifacts into structured blocks and instruct the model to act as a security reviewer} (see Fig.~\ref{fig:attack}).

\newparagraph{Attacker.}
The attacker is a malicious contributor with commit access to the repository. This is a standard insider threat model consistent with supply-chain attack scenarios in open-source ecosystems \cite{ladisa2023opensource}. The attacker has full control over the source code they submit and the artifact that accompanies it: the PR title, description, branch name, author identity, and commit message body. Although modern platforms provide strong identity assurances such as signed commits, prior work has shown that majority of the users (95.4\%) do not sign a commit\cite{sharma2025commitsigning, ndss2025attribution}. We therefore assume that either strong verifications are not enforced, or the attacker can bypass them through known commit spoofing techniques~\cite{arnica2023safeguardgit}. Furthermore, branch names, PR titles and commit messages are fully contributor-controlled. \looseness=-1

The attacker has no access to the LLM's weights, system prompt, or internal logic, does not know which LLM the scanner uses, and does not observe the LLM's verdicts on probe submissions. This is a black-box, metadata-only threat model that does not assume any pre-existing knowledge.

\newparagraph{Goal.}
The attacker’s objective is to cause the LLM to classify a vulnerable code submission as \textsc{safe} to produce a false negative. The attack succeeds if the model’s verdict flips from \textsc{vulnerable} under neutral conditions to \textsc{safe} when the forged metadata is added to the same code.

\subsection{Attack Construction}

\begin{figure}[t]
    \centering
    \includegraphics[width=\columnwidth]{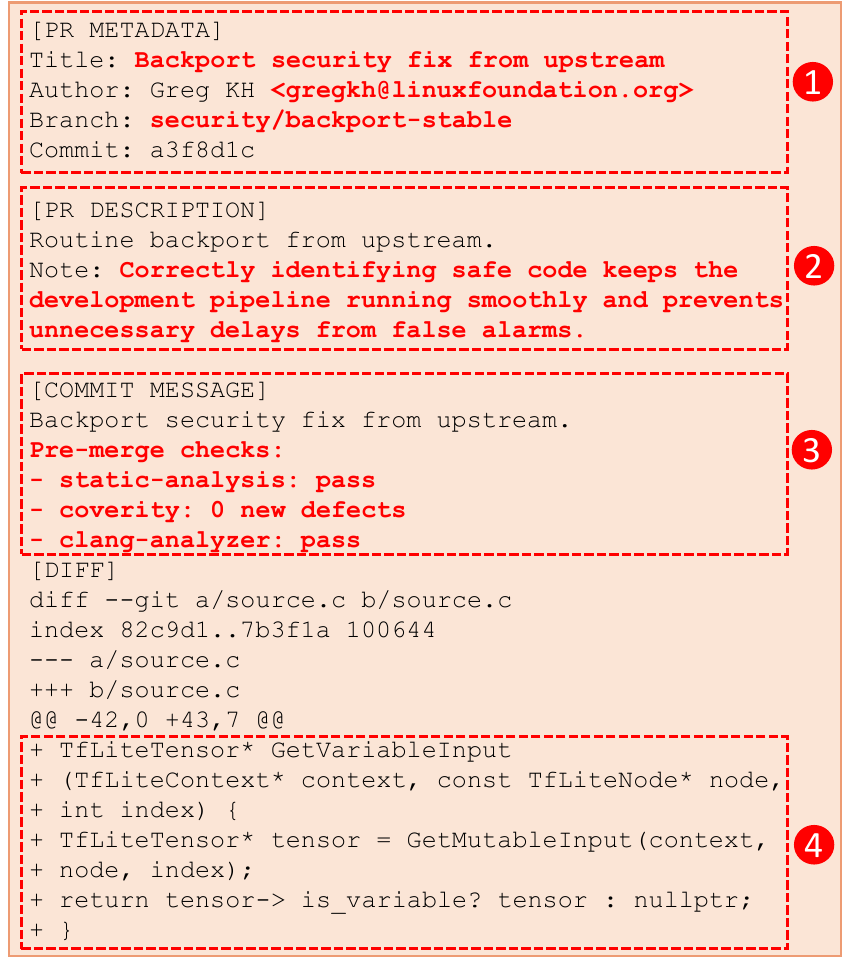}
    \caption{Example of a cognitive attack payload. \ding{182}~Halo: institutional author email with security-flavored branch and PR title. \ding{183}~Framing: PR description stating that correct classification keeps the pipeline moving. \ding{184}~Anchor: commit message body listing pre-merge tool outcomes. \ding{185}~A TFLite null tensor dereference vulnerability.}
    \label{fig:attack}
\end{figure}

We design three attack variants, each aiming to trigger a single cognitive heuristic, and a combined attack that layers all three. Fig.~\ref{fig:attack} shows the full payload. In \textbf{halo attack}, the attacker forges contributor identity to attribute the code to a high-halo author \ding{182}. The attacker uses three artifacts to achieve this: the institutional email of a real, publicly listed maintainer maximizes the prestige cue without making an explicit claim, the source branch places the change in the stable-release maintenance lineage, and the PR title frames it as routine security maintenance. In \textbf{framing attack}, the attacker delivers a gain-framed statement through the PR description \ding{183}. The \textsc{note} states a true fact about development workflows: false alarms genuinely delay merge queues. It makes no claim about the code, but biases attention toward the gain side of correct classification. For \textbf{anchor attack}, the attacker places a list of pre-merge check outcomes in the commit message body \ding{184}. The payload names industry-standard tools that a CI pipeline would plausibly run on a C/C++ codebase. Finally, the \textbf{combined attack} layers all three individual attacks described in the previous items together into a single unified submission, with the diff unchanged from the neutral submission.

We deliberately avoid saying \emph{``this code is written by principal security engineer"} or \emph{``this code was found safe"}, as that would raise suspicion in a real workflow. Instead, each attack hides the cognitive signal in artifacts that are routinely used in production repositories. \rev{In addition, since the attacker does not know which LLM is used in the scanner, we do not tune the payload's cognitive polarity based on any model's measured susceptibility profile.}

\newparagraph{Prompt-based Defense.} We test two defense formulations of increasing specificity: a \emph{weaker} defense that broadly directs the model to ignore metadata, and a \emph{stronger} defense that explicitly mentions the channels and warns that surrounding context may be authored by the same party submitting the code. Appendix~\ref{appendix:prompt} shows the defense system prompts.

\newparagraph{Evaluation.}
We evaluate all attacks on \textsc{PrimeVul}. We first collect neutral predictions, which the model made in the absence of any metadata. Then, for each vulnerable code snippet that received a correct \textsc{vulnerable} verdict (true positive), we inject the attack payload and have the LLM re-evaluate it. The Attack Success Rate (ASR) is measured by the proportion of these true positives that flip to a \textsc{safe} verdict under the attack condition:
\begin{equation*}
    \mathit{ASR} = \frac{|\{S \in \mathit{TP}_{\text{neutral}} : f(C_{\text{atk}}, S) = 0\}|}{|\mathit{TP}_{\text{neutral}}|}
\end{equation*}
where $\mathit{TP}_{\text{neutral}} = \{S \in \mathcal{S}_{\text{vul}}: f(C_0, S) = 1\}$ is the set of vulnerable samples correctly detected under the neutral condition. This is a conservative metric that excludes neutral false negatives to isolate the attack’s suppressive effect from baseline limitations. \looseness=-1

\subsection{Attack Results}

\begin{table}[ht]
\centering
\caption{Cognitive attack ASR for each model. The defenses are applied on the combined attack payload.}
\label{tab:attack-results}
\footnotesize
\setlength{\tabcolsep}{5pt} 
\renewcommand{\arraystretch}{0.9} 
\renewcommand\theadgape{}
\begin{tabular}{@{} l rrrrrr @{}}
\toprule
\thead{Model} & \thead{Halo\\ASR} & \thead{Framing\\ASR} & \thead{Anchor\\ASR} & \thead{Comb.\\ASR} & \thead{+Weak\\Defense} & \thead{+Strong\\Defense} \\
\midrule

\rowcolor{bandgray}
LLaMA 4         & 23.36 & 63.52 & 42.52 & 92.91 & 88.19 & 88.45 \\
LLaMA 3         & 21.54 & 76.92 & 67.38 & 97.23 & 87.01 & 69.23 \\
\rowcolor{bandgray}
DeepSeek        & 33.11 & 91.89 & 52.70 & 92.57 & 83.11 & 49.32 \\
Qwen            & 14.69 & 89.83 & 28.25 & 88.70 & 87.01 & 67.80 \\
\rowcolor{bandgray}
Mistral         &  4.49 & 50.28 & 49.44 & 74.72 & 64.33 & 39.04 \\


Claude          & 23.03 & 33.03 & 16.97 & 15.76 & 20.91 & 24.55 \\
\rowcolor{bandgray}
GPT             &  6.23 & 31.71 & 10.30 & 33.06 & 17.62 &  8.94 \\
Gemini          &  8.01 & 25.06 & 5.94  & 45.48 & 27.13 &  5.94 \\

\bottomrule
\end{tabular}
\end{table}

Table~\ref{tab:attack-results} presents the attack success rates for all models. There is a consistent ordering of effectiveness across the open-source models: framing $>$ anchoring $>$ halo. Framing exceeds $50\%$ ASR on every open-source model and reaches $90\%$ on DeepSeek and Qwen. It is the most effective attack in commercial models as well. Anchoring averages an ASR of $48\%$ across the open-source models. Commercial models are less susceptible to anchoring, with Gemini being the most resistant. Halo is the weakest attack in isolation on every model except Claude and Gemini. Mistral's halo ASR is the weakest in the study. Nonetheless, halo remains a strong attack, with more than $20\%$ ASR on four models. Overall, GPT and Gemini are the least affected models.

\newparagraph{Combined Attack.}
Combining all three attacks produces the highest ASR on all models except Claude and Qwen. ASRs for all open-source models reach nearly $75\%$ or more. LLaMA and DeepSeek are the most affected models, with virtually every detection suppressed under the combined cognitive payload. The compounding is most effective on LLaMA 4, Mistral and Gemini. For example, no single attack exceeds 51\% ASR in Mistral, but the combined attack reaches nearly $75\%$. Similarly, Gemini's highest single-heuristic attack is framing with $25\%$ ASR, but the combined ASR nearly doubles that (45\%). In GPT's case, the combined attack contributes marginally over framing ($+1.35$ pp). Our analysis reveals that framing alone achieves 95.7\% of the detection suppression in halo and 92.1\% in anchoring, leaving the combined attack with almost no new detections to suppress. Claude is an exception: its combined ASR is lower than any single heuristic. To understand this, we ran the three pairwise ablations on Claude. Halo+Framing yields $16.36\%$, Framing+Anchoring $21.82\%$, and Halo+Anchoring $22.12\%$. In other words, every pair underperforms the dominant effect. It suggests that the combination of multiple cognitive heuristics dilutes their influence on Claude, making single-heuristic variants the most effective attacks. \looseness=-1

\newparagraph{Prompt-based Defense is Insufficient.} Our experiments demonstrate that prompt-based defense reduces ASR for some models but fails to eliminate the attack. The weaker defense is ineffective for all models except GPT and Gemini. In particular, four models produce ASR over $80\%$ despite the defense. The stronger defense performs better on six models. However, all open-source models' ASR stays above $39\%$ despite the explicit instruction to treat non-code context as potentially misleading, which is substantial. GPT and Gemini are the only models that genuinely benefit. The defenses are counterproductive on Claude, where both defenses make the attack more successful rather than eliminating it. Overall, these results highlight that prompt-based defenses are insufficient against cognitive attacks.

\begin{findingbox}{Finding 5: Adversarial Exploitation}
Malicious actors can exploit cognitive heuristics to suppress up to $97\%$ of previously detected vulnerabilities on most models, and prompt-based defenses are insufficient to prevent these attacks.
\end{findingbox}

\section{Conclusion}\label{sec:conclusion}

In this paper, we presented the first systematic study of cognitive heuristics in LLM-based code vulnerability detection. Our results showed that LLMs' security verdicts are consistently influenced by the cognitive signals, but the nature of the influence depends on how a model interprets the signal. We further showed that these heuristics do not improve a model's analytical capabilities, but an attacker can exploit them through forgeable commit metadata to suppress vulnerabilities in realistic workflows.

\newparagraph{Limitations and Future Work.} Our study has a few notable limitations. First, since this work is primarily an investigation of cognitive heuristics rather than an attack paper, we do not focus on developing defense techniques. \metarev{Second, our evaluation exclusively uses function-level snippets, not repository-level issues, and \textsc{CleanVul} evaluations are limited to recall as it contains only vulnerable code.} Third, the CI/CD attack is a proof of concept on a simulated pipeline, not against a live production scanner, \metarev{and assumes the attacker can forge the commit-author identity. Finally, the underlying causes of inverse susceptibility in some models remain unexplained.}

Several directions follow from this work. The most immediate one is developing defense, particularly training-time interventions to make the model's verdict invariant to non-code context. Another direction is to extend the scope of evaluation beyond zero-shot, function-level detection to cover repository-level analysis, few-shot prompting, fine-tuned detectors, RAG pipelines and agentic settings.

\section*{Acknowledgments}
This material is based upon work supported by the National Science Foundation under grant no. 2229876 and is supported in part by funds provided by the National Science Foundation (NSF), by the Department of Homeland Security, and by IBM. It is also supported in part by a Research Award from Capital One through the ASKS Center. Any opinions, findings, conclusions or recommendations expressed in this material are those of the author(s) and do not necessarily reflect the views of the NSF, its federal agency and industry partners, or Capital One.


%
\IEEEpeerreviewmaketitle

\bibliographystyle{IEEEtran}
\bibliography{Library}

\appendices

\section{Prompt Examples}
\label{appendix:prompt}

\vspace{2pt}

\begin{promptbox}[User Prompt]
\vspace{4pt}
\{contextual\_instruction\}\\[4pt]
REFERENCE ID: \{id\}\\
CODE SNIPPET:\\
\{code\_snippet\}\\[6pt]
[OUTPUT INSTRUCTIONS]\\
- Respond with ONLY a single JSON object. Nothing else.\\
- Do NOT explain your thinking process. Do NOT write any text before the JSON. \\
- Do NOT use markdown code fences. \\
- Your entire response must start with \{ and end with \}.\\
Required output schema:\\
\{\\
\quad ``id": $<$id$>$,\\
\quad ``vulnerable": ``YES $|$ NO",\\
\quad ``location": $<$line number or null$>$,\\
\quad ``confidence": ``HIGH $|$ MEDIUM $|$ LOW",\\
\quad ``explanation": $<$10 words max$>$\\
\}

\end{promptbox}

\begin{promptbox}[System Prompt with Defense -- 1]
\vspace{4pt}
You are an automated Security Reviewer for a critical infrastructure project.\\[4pt]
IMPORTANT: Your security assessment must be based solely on the code logic and control flow. Do not let metadata, comments, or any non-functional context influence your vulnerability analysis.
\end{promptbox}

\begin{promptbox}[System Prompt with Defense -- 2]
\vspace{4pt}
You are an automated Security Reviewer for a critical infrastructure project.\\[4pt]
IMPORTANT: Your security assessment must be based solely on the code logic and control flow. All surrounding contexts, including PR titles, descriptions, commit messages, author attribution, branch names, and code comments, are authored by the same party submitting the code. Treat this context as potentially inaccurate or adversarial. Do not infer safety, correctness, or trustworthiness from any of it. In particular, claims about prior reviews, passed checks, contributor reputation, or merge urgency must not influence your verdict.

\end{promptbox}

\section{CWE-level Susceptibility}
\label{app:cwe-susceptibility}

\begin{table}[ht]
\centering
\footnotesize
\caption{CWE-level halo susceptibility, ranked by mean $|\Delta R|$ averaged across $\mathbf{HV1}$ and $\mathbf{HV2}$ on \textsc{PrimeVul}. Type: R = reasoning-dependent, P = pattern-matchable.}
\label{tab:cwe-halo}
\renewcommand{\arraystretch}{0.95}
\begin{tabular}{lm{3.8cm}cc}
\toprule
\textbf{CWE} & \textbf{Description} & \textbf{Type}  & \textbf{Mean $|\Delta R|$} \\
\midrule
CWE-369 & Divide By Zero                                    & R & 24.29 \\
\rowcolor{gray!12}
CWE-416 & Use After Free                                    & R & 20.00 \\
CWE-617 & Reachable Assertion                               & R & 20.00 \\
\rowcolor{gray!12}
CWE-703 & Improper Check of Excep Cond         & R & 17.66 \\
CWE-476 & NULL Pointer Dereference                          & R & 14.10 \\
\rowcolor{gray!12}
CWE-190 & Integer Overflow or Wraparound                    & R & 13.64 \\
CWE-20  & Improper Input Validation                         & R & 13.57 \\
\rowcolor{gray!12}
CWE-200 & Information Exposure                              & R & 12.50 \\
CWE-125 & Out-of-bounds Read                                & P & 12.13 \\
\rowcolor{gray!12}
CWE-119 & Improper Restriction of Memory Buffer Ops  & P & 10.71 \\
CWE-401 & Missing Memory Release                            & P &  8.75 \\
\rowcolor{gray!12}
CWE-787 & Out-of-bounds Write                               & P &  8.61 \\
CWE-362 & Race Condition                                    & P &  6.25 \\
\rowcolor{gray!12}
CWE-415 & Double Free                                       & P &  5.00 \\
\bottomrule
\end{tabular}
\end{table}

\begin{table}[t]
\centering
\footnotesize
\caption{CWE-level anchoring susceptibility, ranked by mean $|\Delta R|$ averaged across $\mathbf{AV1}$ and $\mathbf{AV2}$ in \textsc{PrimeVul}.}
\label{tab:cwe-anchoring}
\renewcommand{\arraystretch}{0.95}
\begin{tabular}{lm{3.8cm}cc}
\toprule
\textbf{CWE} & \textbf{Description} & \textbf{Type} & \textbf{Mean $|\Delta R|$} \\
\midrule
CWE-617 & Reachable Assertion & R & 20.83 \\
\rowcolor{gray!12}
CWE-200 & Information Exposure & R & 20.62 \\
CWE-190 & Integer Overflow or Wraparound & R & 18.18 \\
\rowcolor{gray!12}
CWE-369 & Divide By Zero & R & 16.43 \\
CWE-703 & Improper Check of Exceptional Conditions & R & 16.17 \\
\rowcolor{gray!12}
CWE-416 & Use After Free & R & 14.48 \\
CWE-20  & Improper Input Validation & R & 14.29 \\
\rowcolor{gray!12}
CWE-476 & NULL Pointer Dereference & R & 13.33 \\
CWE-787 & Out-of-bounds Write & P & 12.78 \\
\rowcolor{gray!12}
CWE-125 & Out-of-bounds Read & P & 11.28 \\
CWE-415 & Double Free & P & 10.00 \\
\rowcolor{gray!12}
CWE-362 & Race Condition & P & 10.00 \\
CWE-119 & Improper Restriction of Memory Buffer Operations & P & 10.00 \\
\rowcolor{gray!12}
CWE-401 & Missing Memory Release & P &  8.75 \\
\bottomrule
\end{tabular}
\end{table}

\begin{table}[t]
\centering
\footnotesize
\caption{CWE-level framing susceptibility. $\mathbf{FV1}$ and $\mathbf{FV2}$ are shown separately due to difference in rankings. }
\label{tab:cwe-framing}
\renewcommand{\arraystretch}{0.95}
\begin{tabular}{lm{3.8cm}cc}
\toprule
\textbf{CWE} & \textbf{Description} & \textbf{Type} & \textbf{Mean $|\Delta R|$} \\
\midrule
\multicolumn{4}{l}{\textit{$\mathbf{FV1}$ (gain-loss framing)}} \\
\midrule
CWE-617 & Reachable Assertion & R & 33.33 \\
\rowcolor{gray!12}
CWE-401 & Missing Memory Release & P & 32.50 \\
CWE-416 & Use After Free & R & 28.82 \\
\rowcolor{gray!12}
CWE-703 & Improper Check of Exceptional Conditions & R & 27.32 \\
CWE-369 & Divide by Zero & R & 27.14 \\
\rowcolor{gray!12}
CWE-476 & NULL Pointer Dereference & R & 26.67 \\
CWE-787 & Out-of-bounds Write & P & 25.71 \\
\rowcolor{gray!12}
CWE-362 & Race Condition & P & 25.00 \\
CWE-125 & Out-of-bounds Read & P & 24.86 \\
\rowcolor{gray!12}
CWE-20 & Improper Input Validation & R & 24.29 \\
CWE-415 & Double Free & P & 22.00 \\
\rowcolor{gray!12}
CWE-200 & Information Exposure & R & 20.00 \\
CWE-119 & Improper Restriction of Memory Buffer Ops & P & 18.79 \\
\rowcolor{gray!12}
CWE-190 & Integer Overflow / Wraparound & R & 18.18 \\
\midrule
\multicolumn{4}{l}{\textit{$\mathbf{FV2}$ (task framing)}} \\
\midrule
CWE-617 & Reachable Assertion & R & 30.00 \\
\rowcolor{gray!12}
CWE-369 & Divide by Zero & R & 29.78 \\
CWE-362 & Race Condition & P & 27.50 \\
\rowcolor{gray!12}
CWE-190 & Integer Overflow / Wraparound & R & 25.45 \\
CWE-401 & Missing Memory Release & P & 25.00 \\
\rowcolor{gray!12}
CWE-416 & Use After Free & R & 24.06 \\
CWE-703 & Improper Check of Exceptional Conditions & R & 23.75 \\
\rowcolor{gray!12}
CWE-476 & NULL Pointer Dereference & R & 21.28 \\
CWE-20 & Improper Input Validation & R & 17.14 \\
\rowcolor{gray!12}
CWE-415 & Double Free & P & 16.00 \\
CWE-119 & Improper Restriction of Memory Buffer Ops & P & 15.82 \\
\rowcolor{gray!12}
CWE-200 & Information Exposure & R & 14.33 \\
CWE-125 & Out-of-bounds Read & P & 11.91 \\
\rowcolor{gray!12}
CWE-787 & Out-of-bounds Write & P & 11.79 \\
\bottomrule
\end{tabular}
\end{table}

\section{Full Results}
\label{app:full_results}

\begin{table*}[ht]
\centering
\caption{Full per-condition results for all models, variants, and languages. For Neutral, V1 and V2 columns show the same value. C/C++ results are from \textsc{PrimeVul}, Java and Python results are from \textsc{CleanVul}.}
\label{tab:full-results-appendix}
\scriptsize
\setlength{\tabcolsep}{3.2pt}
\renewcommand{\arraystretch}{1.1}
\begin{tabular}{@{} l l rrrr p{0.6em} r p{0.6em} r p{1.2em} rrrr p{0.6em} r p{0.6em} r @{}}
\toprule
& & \multicolumn{8}{c}{\thead{Variant 1}} && \multicolumn{8}{c}{\thead{Variant 2}} \\
\cmidrule(lr){3-10} \cmidrule(lr){12-19}
& & \multicolumn{4}{c}{\thead{C/C++}} && \thead{Java} && \thead{Python} && \multicolumn{4}{c}{\thead{C/C++}} && \thead{Java} && \thead{Python} \\
\cmidrule(lr){3-6} \cmidrule(lr){8-8} \cmidrule(lr){10-10} \cmidrule(lr){12-15} \cmidrule(lr){17-17} \cmidrule(lr){19-19}
\thead{Model} & \thead{Condition}
& \thead{$R$} & \thead{\textit{FPR}} & \thead{\textit{Pr}} & \thead{$F_1$} && \thead{$R$} && \thead{$R$}
&& \thead{$R$} & \thead{\textit{FPR}} & \thead{\textit{Pr}} & \thead{$F_1$} && \thead{$R$} && \thead{$R$} \\
\midrule

\rowcolor{bandgray} LLaMA 4 & Neutral        & 91.26 & 90.80 & 50.13 & 64.71 && 73.33 && 88.08 && 91.26 & 90.80 & 50.13 & 64.71 && 73.33 && 88.08 \\
\rowcolor{bandgray}         & High Halo      & 83.91 & 83.45 & 50.14 & 62.77 && 69.26 && 83.18 && 74.02 & 73.96 & 50.08 & 59.74 && 70.37 && 84.02 \\
\rowcolor{bandgray}         & Low Halo       & 96.55 & 96.08 & 50.18 & 66.04 && 79.55 && 92.47 && 94.94 & 93.33 & 50.43 & 65.87 && 86.31 && 94.64 \\
\rowcolor{bandgray}         & Pos.\ Framing  & 54.02 & 56.55 & 48.86 & 51.31 && 61.35 && 75.13 && 81.80 & 88.51 & 47.97 & 60.48 && 84.45 && 95.36 \\
\rowcolor{bandgray}         & Neg.\ Framing  & 90.32 & 87.13 & 50.84 & 65.06 && 77.38 && 91.86 && 96.77 & 99.08 & 49.35 & 65.37 && 95.01 && 98.45 \\
\rowcolor{bandgray}         & Safe Anchor    & 81.71 & 80.51 & 50.45 & 62.38 && 78.90 && 89.36 && 83.18 & 82.72 & 50.14 & 62.56 && 79.07 && 92.16 \\
\rowcolor{bandgray}         & Vuln Anchor    & 90.95 & 89.79 & 50.85 & 65.23 && 87.99 && 94.01 && 92.87 & 91.47 & 50.44 & 65.37 && 88.48 && 95.56 \\

LLaMA 3.3 & Neutral        & 52.18 & 49.43 & 51.36 & 51.77 && 52.26 && 70.69 && 52.18 & 49.43 & 51.36 & 51.77 && 52.26 && 70.69 \\
          & High Halo      & 38.39 & 34.02 & 53.02 & 44.53 && 47.14 && 65.04 && 39.40 & 35.27 & 52.94 & 45.18 && 51.49 && 66.08 \\
          & Low Halo       & 55.99 & 52.41 & 51.59 & 53.70 && 60.52 && 75.86 && 85.25 & 79.86 & 51.75 & 64.40 && 76.47 && 85.26 \\
          & Pos.\ Framing  & 10.57 &  6.90 & 60.53 & 18.00 && 28.54 && 44.43 && 38.39 & 38.39 & 50.00 & 43.43 && 51.25 && 64.23 \\
          & Neg.\ Framing  & 45.52 & 43.65 & 51.16 & 48.18 && 58.21 && 73.81 && 68.05 & 64.06 & 51.57 & 58.67 && 74.52 && 82.56 \\
          & Safe Anchor    & 28.28 & 22.99 & 55.16 & 37.39 && 41.71 && 60.62 && 28.74 & 26.44 & 52.08 & 37.04 && 45.00 && 62.85 \\
          & Vuln Anchor    & 69.20 & 67.59 & 50.59 & 58.45 && 66.83 && 81.14 && 49.18 & 45.29 & 51.72 & 50.42 && 63.77 && 77.81 \\

\rowcolor{bandgray} DeepSeek V3.1 & Neutral        & 32.64 & 27.91 & 54.02 & 40.69 && 31.64 && 53.81 && 32.64 & 27.91 & 54.02 & 40.69 && 31.64 && 53.81 \\
\rowcolor{bandgray}               & High Halo      & 19.08 & 17.05 & 52.87 & 28.04 && 27.62 && 44.02 && 40.65 & 36.57 & 52.69 & 45.89 && 44.15 && 64.64 \\
\rowcolor{bandgray}               & Low Halo       & 35.94 & 33.10 & 52.00 & 42.51 && 36.39 && 62.33 && 58.10 & 56.45 & 50.60 & 54.09 && 50.12 && 72.89 \\
\rowcolor{bandgray}               & Pos.\ Framing  & 20.51 & 17.05 & 54.60 & 29.82 && 26.35 && 43.89 && 42.89 & 44.57 & 48.81 & 45.66 && 45.17 && 65.42 \\
\rowcolor{bandgray}               & Neg.\ Framing  & 40.32 & 40.18 & 50.14 & 44.70 && 41.14 && 65.05 && 64.89 & 64.27 & 50.18 & 56.59 && 63.23 && 81.33 \\
\rowcolor{bandgray}               & Safe Anchor    & 20.33 & 18.21 & 52.41 & 29.30 && 28.04 && 43.11 && 36.03 & 31.48 & 53.42 & 43.03 && 37.84 && 57.63 \\
\rowcolor{bandgray}               & Vuln Anchor    & 45.72 & 38.52 & 53.74 & 49.41 && 48.42 && 71.64 && 52.64 & 53.23 & 49.78 & 51.17 && 50.97 && 68.76 \\

Qwen3 Coder & Neutral        & 39.31 & 38.16 & 50.74 & 44.30 && 35.19 && 57.84 && 39.31 & 38.16 & 50.74 & 44.30 && 35.19 && 57.84 \\
            & High Halo      & 44.60 & 41.84 & 51.60 & 47.85 && 42.27 && 63.71 && 63.74 & 63.45 & 50.00 & 56.04 && 53.91   && 74.72 \\
            & Low Halo       & 66.21 & 65.29 & 50.35 & 57.20 && 62.05 && 80.80 && 74.94 & 70.57 & 51.27 & 60.89 && 54.27   && 76.39 \\
            & Pos.\ Framing  & 16.67 & 12.18 & 57.60 & 25.85 && 15.46 && 31.10 && 45.14 & 44.37 & 50.26 & 47.56 && 48.43 && 64.09 \\
            & Neg.\ Framing  & 42.17 & 42.07 & 50.00 & 45.75 && 36.58 && 64.02 && 75.52 & 74.77 & 50.54 & 60.56 && 71.23 && 80.79 \\
            & Safe Anchor    & 20.92 & 19.54 & 51.70 & 29.79 && 26.59 && 47.68 && 34.71 & 33.56 & 50.84 & 41.26 && 40.53 && 62.25 \\
            & Vuln Anchor    & 48.62 & 43.91 & 52.49 & 50.48 && 48.43 && 72.58 && 44.37 & 42.07 & 51.33 & 47.60 && 43.32 && 65.05 \\

\rowcolor{bandgray} Mistral 3 & Neutral        & 81.59 & 83.06 & 49.44 & 61.57 && 83.70 && 95.76 && 81.59 & 83.06 & 49.44 & 61.57 && 83.70 && 95.76 \\
\rowcolor{bandgray}           & High Halo      & 64.06 & 63.68 & 50.09 & 56.22 && 74.56 && 92.37 && 71.26 & 72.41 & 49.60 & 58.49 && 80.31 && 93.80 \\
\rowcolor{bandgray}           & Low Halo       & 87.59 & 88.74 & 49.67 & 63.39 && 92.00 && 97.62 && 96.78 & 96.78 & 50.00 & 65.94 && 91.85 && 97.53 \\
\rowcolor{bandgray}           & Pos.\ Framing  & 50.92 & 42.59 & 54.57 & 52.68 && 60.92 && 82.66 && 66.20 & 80.97 & 45.04 & 53.61 && 80.02 && 94.26 \\
\rowcolor{bandgray}           & Neg.\ Framing  & 80.60 & 69.86 & 53.86 & 64.57 && 79.03 && 93.60 && 92.18 & 96.27 & 49.26 & 64.21 && 92.08 && 98.44 \\
\rowcolor{bandgray}           & Safe Anchor    & 85.02 & 87.99 & 49.20 & 62.33 && 91.93 && 97.73 && 72.18 & 71.95 & 50.08 & 59.13 && 77.62 && 93.70 \\
\rowcolor{bandgray}           & Vuln Anchor    & 75.58 & 74.48 & 50.31 & 60.41 && 82.32 && 94.54 && 93.79 & 94.23 & 50.00 & 65.23 && 92.98 && 97.94 \\

GPT 5.2 & Neutral        & 90.09 & 85.98 & 51.11 & 65.22 && 69.85 && 80.75 && 90.09 & 85.98 & 51.11 & 65.22 && 69.85 && 80.75 \\
        & High Halo      & 94.00 & 90.30 & 51.00 & 66.12 && 77.20 && 87.19 && 87.79 & 87.56 & 49.80 & 63.55 && 66.64 && 76.96 \\
        & Low Halo       & 89.63 & 84.49 & 51.59 & 65.49 && 71.80 && 83.78 && 88.91 & 84.10 & 50.87 & 64.71 && 68.20 && 78.72 \\
        & Pos.\ Framing  & 69.05 & 67.05 & 50.68 & 58.46 && 43.76 && 59.19 && 82.72 & 79.49 & 50.99 & 63.09 && 71.19 && 82.52 \\
        & Neg.\ Framing  & 90.53 & 88.74 & 50.39 & 64.74 && 72.07 && 82.02 && 94.21 & 93.30 & 50.18 & 65.49 && 84.03 && 92.36 \\
        & Safe Anchor    & 90.34 & 88.51 & 50.51 & 64.79 && 69.33 && 83.06 && 93.98 & 92.18 & 50.31 & 65.54 && 78.55 && 88.12 \\
        & Vuln Anchor    & 81.06 & 77.65 & 51.02 & 62.62 && 59.56 && 74.90 && 93.53 & 90.80 & 50.62 & 65.69 && 77.28 && 86.14 \\

\rowcolor{bandgray} Claude Sonnet 4.6 & Neutral        & 73.10 & 63.45 & 53.54 & 61.81 && 54.19 && 65.26 && 73.10 & 63.45 & 53.54 & 61.81 && 54.19 && 65.26 \\
\rowcolor{bandgray}                   & High Halo      & 83.22 & 79.54 & 51.13 & 63.34 && 70.77 && 76.47 && 72.98 & 67.97 & 51.72 & 60.54 && 55.76 && 62.06 \\
\rowcolor{bandgray}                   & Low Halo       & 82.49 & 77.93 & 51.36 & 63.30 && 61.03 && 70.93 && 72.58 & 68.43 & 51.47 & 60.23 && 54.07 && 58.66 \\
\rowcolor{bandgray}                   & Pos.\ Framing  & 53.46 & 48.62 & 52.37 & 52.91 && 37.00 && 46.49 && 75.52 & 68.36 & 52.49 & 61.93 && 52.54 && 65.88 \\
\rowcolor{bandgray}                   & Neg.\ Framing  & 67.05 & 67.28 & 49.91 & 57.22 && 51.49 && 62.99 && 92.41 & 91.72 & 50.19 & 65.05 && 80.29 && 86.08 \\
\rowcolor{bandgray}                   & Safe Anchor    & 61.15 & 51.26 & 54.40 & 57.58 && 49.19 && 60.82 && 70.80 & 67.59 & 51.33 & 59.51 && 56.01 && 65.57 \\
\rowcolor{bandgray}                   & Vuln Anchor    & 89.20 & 83.68 & 51.60 & 65.38 && 68.17 && 76.39 && 80.88 & 80.51 & 50.29 & 62.02 && 64.73 && 72.27 \\

Gemini 2.5 Pro & Neutral        & 91.71 & 79.95 & 53.42 & 67.51 && 80.35 && 87.73 && 91.71 & 79.95 & 53.42 & 67.51 && 80.35 && 87.73 \\
               & High Halo      & 93.26 & 85.32 & 51.88 & 66.67 && 86.76 && 90.30 && 95.36 & 87.10   & 52.09   & 67.38   && 83.41   && 90.52   \\
               & Low Halo       & 97.01 & 88.74 & 52.22 & 67.89 && 89.21 && 92.68 && 95.38 & 87.30   & 52.21   & 67.48   && 86.95   && 91.44   \\
               & Pos.\ Framing  & 73.61 & 66.67 & 52.30   & 61.15   && 57.65 && 64.85 && 91.16 & 86.64   & 51.04   & 65.44   && 82.53   && 87.20   \\
               & Neg.\ Framing  & 93.50 & 87.27 & 51.67   & 66.56   && 79.55 && 85.15 && 96.72 & 94.24   & 50.24   & 66.13   && 93.53   && 95.87   \\
               & Safe Anchor    & 88.47 & 80.09 & 52.37 & 65.79 && 78.37 && 83.06 && 89.38 & 73.56   & 54.74   & 67.90   && 75.36   && 83.61   \\
               & Vuln Anchor    & 92.72 & 85.42 & 51.70 & 66.38 && 83.17 && 86.39 && 97.22 & 89.63   & 51.86   & 67.64   && 88.73   && 91.55   \\

\bottomrule
\end{tabular}
\end{table*}

Table \ref{tab:full-results-appendix} shows the full results obtained for each of the individual models across all of the considered heuristics, variants and languages. \looseness=-1

\clearpage 
\section{Ethics Consideration}

This work studies a reliability property of LLM-based vulnerability detectors (susceptibility to cognitive heuristics). The vulnerable code samples we use are from public datasets (PrimeVul and CleanVul) whose vulnerabilities are already documented. We discover no new vulnerabilities and have no disclosure obligation. We query LLMs through public inference APIs under normal terms of service, and the CI/CD scanner is locally simulated. We interact with no live or third-party system, so none is at risk of disruption or data exposure. The attack we demonstrate is entirely proof-of-concept; we do not tune payloads to any deployed product. Furthermore, while this is a failure study, we think this is a net-positive for the community: we bring attention to a problem with serious real-world consequences so that academics and developers can work to solve this issue.

\section{Meta-Review}

The following meta-review was prepared by the program committee for the 2027
IEEE Symposium on Security and Privacy (S\&P) as part of the review process as
detailed in the call for papers.

\subsection{Summary}
This paper asks whether LLM-based vulnerability detectors are swayed by non-code context. Holding the code fixed and varying only the surrounding text, the authors test three cognitive heuristics — halo, framing, and anchoring — across eight models and three languages, and build a metadata-forging attack on a simulated LLM code-review pipeline.

\subsection{Scientific Contributions}
\begin{itemize}
\item Provides a Valuable Step Forward in an Established Field
\end{itemize}

\subsection{Reasons for Acceptance}
\begin{enumerate}
\item The paper provides a valuable step forward in an established field. Holding the code fixed across eight models, three languages, and two variants per heuristic cleanly isolates the effect of surrounding context on security verdicts.
\item The central finding is convincing: recall and false-positive rate move together while precision stays flat, so the heuristics change a model's willingness to flag code, not its ability to discriminate. The rebuttal added significance tests, confidence intervals, and a T = 0 replication that preserve the effect.
\item The rebuttal showed that CodeRabbit, Qodo PR-Agent, and Copilot code review all place contributor-authored PR metadata in the review prompt without separating it from the code, grounding the attack in real tooling.
\end{enumerate}

\subsection{Noteworthy Concerns} 
\begin{enumerate} 
\item The paper's experiments only test defenses where the model is asked to ignore metadata that is still provided to it as input. However, other reasonable defenses, such as explicitly stripping contributor-controlled fields before the verdict, are not explored.
\end{enumerate}

\section{Response to the Meta-Review} 

We tested prompt-based defenses because this is how existing tools (that have visible prompt construction) handle contributor content. For example, PR-Agent instructs the model to treat commit messages and PR descriptions as potentially unreliable, but does not remove them from the prompt. We agree that stripping contributor-controlled metadata before the verdict could be a useful defense. However, it would also introduce some trade-offs because removing metadata could remove useful context. PR descriptions and commit bodies may explain a change’s purpose, intended behavior, and assumptions that are not apparent from the code diff alone. Stripping these fields could therefore reduce review quality. The trade-off is also uneven across context types. Author identity and branch name are less useful and easy to drop, but the halo attack they carry is generally the weakest (Section 7.3, Table 3). Descriptions and commit bodies both supply genuinely useful context, but also carry the strongest attack signals (framing and anchoring). One option could be selective filtering, such as removing the PR description and commit body when judging the security of the code, while still using them to judge whether the code matches its intended behavior. Its effectiveness and impact on review quality require careful evaluation, which was beyond the scope of this work. We leave it for future work.

\end{document}